\documentclass[aps,prx,onecolumn,superscriptaddress,amsmath,nofootinbib,eqsecnum,amssymb,11pt,floatfix]{revtex4-2_custom}

\usepackage{physics}
\usepackage{amssymb,amsmath,mathdots}
\usepackage{mathrsfs}
\usepackage{graphicx}
\usepackage{lmodern}
\usepackage{amsmath}
\usepackage{bm,bbold}
\usepackage{pgfplots}
\pgfplotsset{compat=1.17}
\usepgfplotslibrary{external}
\usepackage[normalem]{ulem}
\usepackage[colorlinks=true]{hyperref}
\hypersetup{
    colorlinks=true,
    linkcolor=blue,
    citecolor=blue,
    urlcolor=blue
} 
\usepackage{color}

\newcommand{\beq} {\begin{equation}}
\newcommand{\eeq} {\end{equation}}
\newcommand{\bea} {\begin{eqnarray}}
\newcommand{\eea} {\end{eqnarray}}
\newcommand{\be} {\begin{equation}}
\newcommand{\ee} {\end{equation}}
\renewcommand{\(}{\left(}
\renewcommand{\)}{\right)}
\renewcommand{\[}{\left[}
\renewcommand{\]}{\right]}
\newcommand{\Langle}{\left\langle}
\newcommand{\Rangle}{\right\rangle}

\newcommand{\non}{\nonumber}

\newcommand{\ve}[1]{{\bm #1}}
\newcommand{\vp}{\ve{p}}

\newcommand{\vx}{\ve{x}}

\newcommand{\vk}{\ve{k}}
\newcommand{\vR}{\ve{R}}

\newcommand{\vK}{\ve{K}}

\DeclareMathOperator{\sgn}{sgn}

\newcommand{\xcentcolon}{%
  \mathrel{\vbox{\hbox{$:$}\kern.2ex}}%
}

\begin{document}

\title {Bosonized theory of de Haas-van Alphen quantum oscillation in Fermi liquids}
\author{Yuxuan Wang}
\email[email address: ]{yuxuan.wang@ufl.edu}
\affiliation{Department of Physics, University of florida, Gainesville, Fermi liquid 32611, USA}
\begin{abstract}
The de Haas-van Alphen effect (dHvA) of a 2d Fermi liquid remains poorly understood, due to the $\sim\mathcal{O}(1)$ contribution to the oscillations of grand potential from the oscillatory part of the fermionic self energy, which has no known closed-form solution.  In this work, we solve this problem via coadjoint-orbit bosonization of the Fermi surface. Compared with the fermionic formalism, the issue of the oscillatory self energy is circumvented.  As an effective field theory, Landau parameters $F_{n}$ directly enter the theory. We use the bosonized theory to derive the energies of cyclotron resonance and specific heat, which are consistent with Fermi liquid theory. Via a mode expansion, we show that the problem of dHvA is reduced to 0+1D quantum mechanics. 
We obtain analytic expressions for the behavior of dHvA at low and high temperatures, which deviate from the well-known Lifshitz-Kosevich formula. We contrast this behavior with that of 3d Fermi liquids, for which we show such deviations are parametrically small. We discuss the effects of disorder on dHvA within the bosonized theory.
\end{abstract}
\date{\today}
\maketitle
\newpage

\tableofcontents

\newpage
\section{Introduction}
In gapless fermionic systems, quantum oscillations in a weak external magnetic field, i.e., de Haas-van Alphen (dHvA) and Shubnikov-de Haas effects, encode important information about the low-energy properties of the system, and have been widely used in experiments to determine the size, shape, and topology of the Fermi surface (FS)~\cite{Shoenberg1984,Mikitik1999,SRO-2000,mikitik-2004, QOgraphite2004,QOzhang2005,taillefer-2007,maslov-hamlin-2014,Sebastian-2015,Sebastian-SmB6-2015,Berry-Rashba,cd3as2-dhva,wu2019anomalous,
ye2019haas,Alexandradinata2023ar,QOkagome2022,QOkagome2023,li2023quantumoscillationstopologicalphases,Guo2024}.

For noninteracting particles, dHvA can be obtained by directly computing the grand potential $\Omega$ by summing the Landau levels. This result is known as the Lifshitz-Kosevich (LK) formula~\cite{LK1956}. In 2d, its oscillatory part is given by 
\be
\Omega_{\rm osc} = N_{\Phi} T \sum_{k=1}^{\infty} \frac{(-)^k \cos(k\mathcal A_{\rm FS}/B + k\gamma )}{k\sinh(2\pi^2 k T/\omega_c)},
\ee
where $N_{\Phi}=BL^2/2\pi$ is the Landau level degeneracy, $T$ the temperature, $\mathcal A_{\rm FS}$  the area enclosed by the FS, $B$  the magnetic field, $\omega_c$  the cyclotron frequency (or the Landau level spacing), and $\gamma$  the Berry phase~\cite{chang-niu-1996,Mikitik1999,Alexandradinata2023ar} of the Bloch wave around the FS\footnote{For simplicity, we will not consider Berry phase effects in this work. Such effects have been treated within bosonization in Ref.~\cite{ye_wang_second} in the noninteracting limit.}. In 3d, a similar LK formula can be obtained by integrating $\Omega_{\rm osc}$ over the momentum along the $B$ field (see Eq.~\eqref{eq:LK3d}).

Beyond the non-interacting limit, the treatment of dHvA within a many-body field-theoretic framework is much less straightforward. As the magnetization contains terms such as $\sim\cos(a/B)$ that have an essential singularity at $B=0$, dHvA cannot be captured  within the linear (or nonlinear) response theory, even if the system is in a weak magnetic field and is weakly interacting. Instead, one has to directly compute the grand potential $\Omega$ with interactions in the Landau level basis~\cite{Luttinger1961,Gorkov-1962}. This approach faces several challenges. First, even in the absence of interactions, the energies of the Landau levels implicitly depend both on the  dispersion  and the  geometry of the Bloch function, and have to be computed outside the field theory using, e.g., semiclassical methods. Second, the fermionic self energy in the presence of a magnetic field, which $\Omega$ depends on, differs from their results at $B=0$ and is oscillatory in $1/B$, i.e.,
$\Sigma(B) = \Sigma_{B=0} + \Sigma_{\rm osc}(B)$. Unlike $\Sigma_{B=0}$, $\Sigma_{\rm osc}(B)$ is generally difficult to evaluate. 
Fortunately, in 3d, due to destructive interference  along the field direction, $\Sigma_{\rm osc}$ is strongly suppressed as $\Sigma_{\rm osc}/\Sigma_{B=0}=\mathcal{O}{\[\({\omega_c}/{E_F}\)^{3/2}\]}$~\cite{Luttinger1961,Gorkov-1962,maslov-2003,Nosov2024}, where $E_F$ is the Fermi energy, and to compute $\Omega_{\rm osc}$ to leading order one can simply use the zero-field self energy. Taking advantage of this suppression, one can show that the dHvA for a 3d Fermi liquid retains the form of the LK formula, except that the cyclotron frequency $\omega_c = B/m$ is defined with $m$ as the effective mass.

However, the situation in 2d is much more complicated. First of all, for $T\ll \omega_c$, fixing chemical potential $\mu$ and particle number $N$ lead to different results, even for a non-interacting Fermi gas. Indeed, in the former case, the chemical potential is almost always (with a unity probability) between Landau levels, while in the latter case, a Landau level is almost always partially filled.  For partially filled Landau levels, interactions are expected to play a much more prominent role, and the ground state is not adiabatically connected to the zero-field Fermi liquid, and can be e.g., charge-ordered~\cite{koulakov1996chargedensity, lilly1999anisotropic, pan1999anisotropic} or topologically ordered~\cite{Stormer1999}. Within this work, we will take $\mu$ as fixed, i.e., the grand canonical emsemble. In experiments, this can be achieved by connecting the system to a large reservoir, e.g., to source and drain contacts used in transport measurements. 

Conspicuously, the dHvA in 2d Fermi liquids remains poorly understood, even with $\mu$ fixed. This is not due to any fundamental reason -- after all, dHvA measures low-energy properties around the Fermi surface, which should be completely captured by the Fermi liquid theory. Instead, the main issue is that for a 2d Fermi liquid in a $B$ field, the oscillatory component in the self energy is less suppressed than in 3d, $\Sigma_{\rm osc}/\Sigma_{B=0}=\mathcal{O}(\omega_c/E_F)$. Such an oscillation contributes to the grand potential $\Omega$ at the same order ($\sim\mathcal{O}\[(\omega_c/E_F)^2\]$) as the oscillation in the non-interacting part of $\Omega$, and cannot be neglected even at weak field~\cite{Miyake1993,CurnoeStamp1998,maslov-2003}. In general, different harmonics of $\Sigma_{\rm osc}(B)$ in $\exp(ik\mathcal{A}_{\rm FS}/B)$ are coupled to each other, and a closed-form solution is unknown. Although at high enough temperatures with $T\gtrsim \omega_c$, these oscillations are exponentially suppressed, they remain important in the dHvA amplitudes $A_{k>1}$, which are exponentially suppressed even more strongly. As a result, only the amplitude of the leading harmonic is known~\cite{maslov-2003} at $T\gtrsim \omega_c$ for a 2d Fermi liquid and follows the LK formula. Similar issues occur in the dHvA for a 3d non-Fermi liquids, and a special solvable case was recently discussed in Ref.~\cite{Nosov2024}.

We show in this paper that the difficulty of the oscillatory self energy can be avoided by using the bosonization approach. Higher-dimensional bosonization of FS's has a long history~\cite{HaldaneBosonization2005,FradkinBosonization1994,FradkinBosonizationPRL, khvesh,Houghton_2000,DDMS2022,kim2023,balents2024,Huang2024,ravid2024electronslostphasespace,ye_wang_second,Delacretaz2025}, but until very recently the issue of magnetic response remained unaddressed by bosonization. Indeed, in earlier theories of bosonization, one first divides the FS into patches, bosonizes the patches, and then assembles them together. In such a theory, the area $\mathcal A_{\rm FS}$ enclosed by the FS, which dHvA crucially relies on, does not enter. Recently, bosonization has been reformulated~\cite{DDMS2022} in terms of the coadjoint orbit of canonical transformations, which describes the deformation of the Fermi sea, resulting in an inherently nonlinear field theory in the \emph{phase space} $(\vx, \vp)$ without resorting to patches. In Refs.~\cite{ye_wang_first,ye_wang_second}, we constructed a bosonic field theory for a 2d Fermi gas in a weak magnetic field. Remarkably, it was found that $\mathcal A_{\rm FS}$ appears explicitly in a topological term, which was neglected in earlier bosonized theories of FS. Proper quantization the theory requires utilizing the magnetic translation symmetry and methods in noncommutative field theory, and it was found that the field in the phase-space space can be decomposed into $N_{\Phi}$ components, each labeled by a different magnetic momentum~\cite{ye_wang_first}. Upon the mode expansion of each field around the FS, Landau-level physics naturally emerges. Focusing on the zero-mode sector, the topological term becomes a $\theta$-term in 0+1D, and was shown to be responsible for dHvA~\cite{ye_wang_first,ye_wang_second}. As an inadvertent advantage, the analysis of dHvA is reduced from a fermionic many-body problem in the fermionic language to a quantum mechanics problem in the bosonized language. 

In Landau's Fermi liquid theory~\cite{baym2008landau}, interaction effects are captured by the Landau parameters $F(\theta,\theta')$ that describe density-density interactions in the forward scattering channel.  The Landau parameters can be straightforwardly included~\cite{DDMS2022} in the bosonized action as a semiclassical interaction (see Eq.~\eqref{eq:1}) which is local in $\vx$ and nonlocal in $\vp$. In this work, we incorporate the Landau parameters into the bosonized theory in the presence of a $B$ field. For our purposes, we convert the basis of the base manifold (phase space) from $(\vx, \vp)$ to the ``eigenbasis" that label the $N_{\Phi}$ modes of bosons. As this eigenbasis is a highly nonlocal one, the Landau parameter term that is local-in-$\vx$  in the new basis becomes an \emph{all-to-all} coupling among the $N_{\Phi}$ modes.

By quantizing the theory in the Fock space of the bosonic excitations, we obtain the energies of the collective modes corresponding to cyclotron resonance. The result is consistent with Kohn's theorem~\cite{Kohn}, which states that the cyclotron frequency is unrenormalized by interaction effects in a Galilean invariant system. The full spectrum of the bosonic excitations allow for evaluation of the specific heat by analogy with black body radiation. Within coadjoint-orbit bosonization, the evaluation of specific heat is difficult due to the UV/IR mixing issue~\cite{DDMS2022}. However, adding a $B$ field resolves this issue by introducing an inherent UV length scale $\ell_B=1/\sqrt{B}$~\cite{ye_wang_first}, and we obtain the correct specific heat for a 2d Fermi liquid. Turning to dHvA, we quantize the theory in the zero-mode sector and evaluate the free energy. We explicitly show that for a 2d Fermi liquid, the behavior of dHvA deviates~\cite{CurnoeStamp1998,maslov-2003} from the LK formula at different temperatures. In particular, the second harmonic has an amplitude $A_2$ which changes sign upon increasing temperature. Our key results are schematically summarized in Fig.~\ref{fig:1}. 

We contrast the 2d result with 3d dHvA, and demonstrate that for the latter case destructive interference along the direction of the field strongly suppresses any deviation from the LK formula by a factor of $\mathcal{O}\[(\omega_c/E_F)^{1/2}\]$. This reduction ratio is consistent with the result from the fermionic theory~\cite{Luttinger1961,Gorkov-1962,maslov-2003,Nosov2024}. Furthermore, our bosonization approach allows us to quantitatively obtain the small corrections to the LK formula, which we derive for weak interactions.

Finally, we consider the effect of disorder on dHvA within bosonization. Instead of analyzing a microscopic model for disorder in which magnetic translation symmetry is broken, we modify the bosonic action according to the fermionic spectral function in the presence of impurity scattering, and obtain the expected Dingle factor, which smears all amplitudes of dHvA.

The remainder of the paper is organized as follows. In Sec.~\ref{sec:intro} we review the bosonization of a FS in the presence of a $B$ field. In Sec.~\ref{sec:LP} we incorporate Landau parameters in the action of $N_{\Phi}$ bosonic fields labeled by distinct magnetic momenta. In Secs.~\ref{sec:kohn} and \ref{sec:ct} we verify consistency with known results for 2d Fermi liquids by calculating the cyclotron resonance energies and specific heat. We present our key results on the dHvA of 2d Fermi liquids in Sec.~\ref{sec:dhva}, and in Sec.~\ref{sec:3d} we contrast it with 3d Fermi liquids. In Sec.~\ref{sec:dis} we model the effect of disorder and compute the Dingle factor. In Sec.~\ref{sec:conc} we present our conclusion.

\section{Review: bosonization of a 2d Fermi gas in a weak magnetic field}\label{sec:intro}

We begin with a brief review of the bosonization of a 2d Fermi gas in a weak magnetic field, which we developed in Ref.~\cite{ye_wang_first} based on the coadjoint-orbit approach in Ref.~\cite{DDMS2022}.  In this work, the ``weak field" is defined via the usual condition 
\be
k_F\ell_B\gg 1 \,\Leftrightarrow\, \frac{B}{k_F^2}\ll 1 \,\Leftrightarrow\, \frac{\omega_c}{E_F} \ll 1,
\ee
where $k_F$ is the Fermi wave vector and $E_F$ is the Fermi energy.

From a coherent-state  path integral~\cite{Luca_talk,ye_wang_first}, one can prove that the bosonized path integral of a 2d Fermi gas is given by
\be
Z = \int \mathcal{D}U(\vx, \vp,t) \exp{i\int\frac{dt d\vx d\vp}{(2\pi)^2}\[f_0(\vx,\vp)\star U^{-1}(\vx,\vp,t)\star i\partial_tU(\vx,\vp,t) - f(\vx,\vp,t)\star\epsilon(\vx,\vp)\]},
\label{eq:action}
\ee
where the phase-space field $U(\vx,\vp)=e^{i\phi(\vx,\vp)}$ is a Lie group element corresponding to the canonical transformation, $\mathcal DU$ is its Haar measure, $f_0(\vx,\vp)$ is the single-particle distribution function of the ground state, $\epsilon(\vx,\vp) = \epsilon_0(\vx,\vp) - \mu$ is the single-particle energy, and
\be
f(\vx,\vp,t)= U(\vx,\vp,t)\star f_0(\vx,\vp)\star U^{-1}(\vx,\vp,t).
\ee
In the above, $\star$ is the Moyal star product~\cite{KamenevBook}:
\begin{equation}
    F(\vx,\vp) \star G(\vx,\vp) = F(\vx,\vp) \exp\left( i \frac{\overleftarrow{\nabla_\ve{x}}\overrightarrow{\nabla_{\ve{p}}}-\overleftarrow{\nabla_{\ve{p}}}\overrightarrow{\nabla_{\ve{x}}}}{2} \right) G (\vx,\vp).
     \label{eq:star}
\end{equation} 
Physically, the Moyal algebra encodes the fact that the phase space is non-commutative, i.e., $[\widehat \vx, \widehat \vp]\neq 0.$ The first term in the action \eqref{eq:action} can be regarded as a Wess-Zumino-Witten term~\cite{DDMS2022}, and the second term is the Hamiltonian of the many-body system $H = \int_{\vx,\vp} f(\vx, \vp) \epsilon(\vx, \vp)$.~\footnote{Note that $\int_{\vx,\vp} A\star B = \int_{\vx,\vp} AB$.}

In a weak magnetic field for which the lattice effect is unimportant ($Ba_0^2<Bk_{F}^{-2}\ll 1$ where $a_0$ is the lattice constant),  the action can be rewritten in the phase space using the basis of guiding center $\vR$ and kinetic momentum $\vk$, where 
\be
\vk = \vp + e\ve{A},~~~\vR = \vx -(\vk\times \hat{\ve{z}})/B.
\label{eq:Rk}
\ee
The action can be rewritten as
\be
S= \int \frac{dt d\vR d\vk}{(2\pi)^2} \[f_0(\vk)\star U^{-1}(\vR,\vk,t)\star i\partial_tU(\vR,\vk,t) - f(\vR,\vk,t)\star\epsilon(\vk)\].
\ee
Moreover, it is straightforward to show that the star product in \eqref{eq:star} can be reexpressed as a direct product of two separate star products:
 \begin{align}
  F (\vR,\vk)\star G(\vR,\vk) &= F(\vR,\vk) \exp \left( i \frac{\overleftarrow{\nabla_\ve{R}}\times\overrightarrow{\nabla_{\ve{R}}}}{2B} \right) \otimes \exp\left( -i  B \frac{ \overleftarrow{\nabla_\ve{k}} \times \overrightarrow{\nabla_{\ve{k}}}}{2} \right) G(\vR,\vk)  \non\\
  &\equiv F(\vR,\vk) \star_{\vR}\star_{\vk} G(\vR,\vk).
 \end{align}

As is commonly done in noncommutative field theory~\cite{Douglas2001rmp},  the integration of Moyal products with $\star_{\vR}$ over $\vR$ can be converted to a trace of operators defined via Weyl quantization, i.e., we have
 \be
\int \frac{d\vR}{2\pi B^{-1}}A(\vR)\star_{\vR} B(\vR) = \Tr[A(\widehat{\vR}) B (\widehat{\vR})],
\ee
where $[\widehat R_{x}, \widehat R_y]=i/B$. As shown in Ref.~\cite{ye_wang_first} via the magnetic translation operators generated by $B\widehat R_{x}$ and $ B\widehat R_y$, the trace can be performed as a sum over $N_{\Phi}$ magnetic Bloch states in the magnetic Brillouin zone (mBZ), where $N_{\Phi} = BL^2/2\pi$ is the Landau level degeneracy:
\be
\int \frac{d\vR}{2\pi B^{-1}}A(\vR)\star_{\vR} B(\vR) = \Tr[A(\widehat{\vR}) B (\widehat{\vR})]= \sum_{\bm{K}\in \rm mBZ}\bra{\bm{K}}A(\widehat{\vR}) B (\widehat{\vR})\ket{\bm{K}}.
\ee
In addition, due to the magnetic translation symmetry of the Hamiltonian, it suffices~\cite{ye_wang_first} to only include in the coherent-state path integral over $\textstyle U(\vR, \vk)$ configurations that are invariant under magnetic translation, i.e., we require
\be
\bra{\bm{K}}U(\widehat\vR, \vk)\ket{\bm{K}'}=U_{\vK}(\vk)\delta_{\vK\vK'}.
\ee

The action then becomes 
\be
S = \sum_{\bm{K}\in \rm mBZ}\int \frac{dt d\vk}{2\pi B} \[f_0(\vk)\star U_{\vK}^{-1}(\vk,t)\star i\partial_t U_{\vK}(\vk,t) - U_{\vK}(\vk,t) \star f_0(\vk)\star U_{\vK}^{-1}(\vk,t)\star\epsilon(\vk)\],
\ee
where we have replaced $\star_{\vk}$ with $\star$, i.e.,
\begin{align}
  F(\vk) \star G(\vk) \equiv F(\vk)  \exp\left( -i  B \frac{ \overleftarrow{\nabla_\ve{k}} \times \overrightarrow{\nabla_{\ve{k}}}}{2} \right) G(\vk).
 \end{align}
 
In a weak magnetic field, $B k_F^{-2} \ll 1$. As ${\nabla_{\vk}}\sim 1/k_F$, the star product $\star_{\vk}$ can be approximated by the leading-order gradient expansion. Let us replace $U_{\vK}$ with $U_i$ for compactness, where $i\in [1,N_{\Phi}]$ accounts for the entire mBZ.
Defining a compact field $\phi_i$ via $U_i(\vk,t) = \exp(i\phi_i(\vk,t))$, it was found in Ref.~\cite{ye_wang_first} that up to quadratic level in $\phi$,
$S =S_{\rm cb}+S_{\rm w}$, with 
\begin{align}
S_{\rm cb}=& \sum_{i=1}^{N_{\Phi}} \int\frac{dt  d\theta}{4\pi }\partial_\theta \phi_{i}\left(\dot\phi_{i}  - \omega_c \partial_\theta \phi_{i} \right)\non \\
S_{\rm w}=& \sum_{i=1}^{N_{\Phi}}  \int\frac{dt  d \vk}{2\pi }f_0(\vk) \[-\dot\phi_i+ \frac 12 (\partial_{\vk}\times \partial_{\vk} \phi_i) \dot{\phi_i}\],
\label{eq:free}
\end{align}
where $\omega_c = B/m$ is the cyclotron frequency, and $m$ is the electron mass. In the path integral Eq.~\eqref{eq:action}, the Haar measure can be rewritten as $\mathcal{D}U(\vx, \vp,t)=\prod_{i=1}^{N_{\Phi}} \mathcal{D}{\phi}_{i}(\vk,t)$.  In $S_{\rm cb}$, the field $\phi_i(\theta,t)$ is taken at the FS, where $\theta$ is the angle parametrizing the FS, and the action describes $N_{\Phi}$ species of chiral bosons propagating along the FS~\cite{FJ1987}. On the other hand, $S_{\rm w}$ is only nonzero when $\phi_i(\vk,t)$ is not-single valued~\cite{ye_wang_first}, due to vorticity in momentum space or winding in time. As we will show in Sec.~\ref{sec:dhva}, it contains topological terms that do not enter the equation of motion, but rather affect the quantization of the theory.

\section{Landau-parameter term in a weak magnetic field}\label{sec:LP}

In Landau's Fermi liquid theory,  interaction effects are effectively described by density-density interaction in the forward scattering channel~\cite{baym2008landau}. We have the effective Hamiltonian
\be
H = \int\frac{d\vx d\vp}{(2\pi)^2} f(\vx, \vp) \epsilon(\vx,\vp) + \int \frac{d\vx d\vp d\vp'}{16\pi^4} \tilde F (\vp, {\vp'})\, \delta f(\vx, \vp) \delta f(\vx, \vp') .
\label{eq:1}
\ee
Here $\epsilon(\vx, \vp)$ is the energy of the \emph{quasiparticle}. For a translationally invariant system, we take $\epsilon(\vp) = v_F(|\vp|-k_F)$ and $v_F= k_F/m$, where $m$ is the \emph{effective mass}. Equivalently, as an effective field theory, we add to the action in Eq.~\eqref{eq:action} an extra term
\be
S_{\rm LP} = -\int \frac{d\vx d\vp d\vp'}{16\pi^4}\tilde F (\vp, {\vp'}) \,\delta f(\vx, \vp) \delta f(\vx, \vp') .
\ee

As we reviewed above, in the bosonized field theory, we have
\begin{align}
\delta f (\vx, \vp,t) = U(\vx, \vp,t)  \star f_0 (\vx, \vp,t) \star U^{-1}(\vx, \vp,t)  - f_0 (\vx, \vp).
\end{align}
In a weak magnetic field, we switch to the $(\vR,\vk)$ coordinates. At leading order~\cite{ye_wang_first},
\be
\delta f (\vR, \vk,t) = B\partial_\theta \phi(\theta_{\vk},t)\, \delta(k-k_F) + \cdots,
\label{eq:df}
\ee
and $S_{\rm LP}$ becomes Gaussian:\footnote{By doing this, we have neglected the damping effects of the quasiparticles due to interactions, which is weak at low energies for Fermi liquids.}
\be
S_{\rm LP} = -\frac{\omega_c}{2}\int \frac{d\vR }{2\pi B^{-1}} \int\frac{d\theta d\theta'}{4\pi^2} \int dt F(\theta, \theta')\, \partial_{\theta}\phi(\vR,\theta, t) \partial_\theta \phi(\vR',\theta' ,t),
\label{eq:LP0}
\ee
where $\omega_c=B/m$ with $m$ being the effective mass. In the above we have defined the Landau parameter as $F(\theta,\theta')= {m}\tilde F(\vk_F,\vk_F')/\pi$ for two Fermi momenta with polar angles at  $\theta$ and $\theta'$, and according to Eq.~\eqref{eq:Rk},
\be
\vR'-\vR= \frac{k_F}{B}\(\sin\theta-\sin\theta', -\cos\theta +\cos\theta'\).
\ee

Like before, the integral over noncommutative $\vR$ coordinates can be rewritten as a trace over Bloch states in the mBZ, and keeping only (magnetic-)translationally invariant modes, we have at quadratic order in $\phi$
\be
S_{\rm LP} = -\frac{\omega_c}2  \int \frac{dt d\theta d\theta'}{4\pi^2} \sum_{\ve{K}\in \textrm{ mBZ}}\bra{\vK} \partial_\theta \phi(\widehat \vR)\ket{\vK} \bra{\vK}\partial_{\theta'}\phi(\widehat \vR') \ket{\vK}  F(\theta - \theta').
\label{eq:5}
\ee
 We note that $\vR'$ and $\vR$ are related $\widehat\vR'=\widehat t(\ve{d}) \widehat \vR \,\widehat t(\ve{d})^{-1}$, where $\ve{d} = k_F/B(\sin\theta - \sin \theta', -\cos\theta + \cos\theta')$, and 
\be
\widehat t(\ve{d})=\exp[- i \frac{d_x \widehat R_y - d_y \widehat R_x}{\ell_B^2}]
\ee
is the magnetic translation operator.
We can thus express the second matrix element in Eq.~\eqref{eq:5} as
\be
\bra{\vK}\partial_{\theta'}\phi(\widehat \vR') \ket{\vK}  = \bra{\vK} \widehat{t}(\ve{d})\partial_{\theta'}\phi(\widehat \vR) \widehat t(\ve{d})^{-1} \ket{\vK}.
\ee
From the properties of the magnetic translation operators~\cite{Haldane2018b}
\be
\widehat t(\ve{d}_1)\widehat t(\ve{d}_2) = \exp[i (\ve{d}_1 \times \ve{d}_2) /\ell_B^2]\widehat t(\ve{d}_2)\widehat t(\ve{d}_1),
\ee
we see that $\widehat t(\ve{d})^{-1} \ket{\vK} = \ket{\vK'(\theta,\theta')}$, 
where the shifted momentum is given by
\be
\vK'(\theta,\theta') = \vK +k_F(\cos\theta - \cos \theta', \sin\theta - \sin\theta') \mod \ve{M}_{x,y},
\label{eq:diff}
\ee
and $\ve{M}_{x,y}$ are the reciprocal basis vectors defining the mBZ, and $|\ve{M}_{x}\times \ve{M}_{y}|=2\pi B$.  As in the free fermion case, we denote $\bra{\vK} \partial_\theta \phi(\widehat \vR)\ket{\vK}$ as $\partial_\theta \phi_\vK$. Then, the second matrix element in Eq.~\eqref{eq:5} is 
\be
\bra{\vK}\partial_{\theta'}\phi(\widehat \vR') \ket{\vK}  
= \bra{\vK'}\partial_{\theta'}\phi(\widehat \vR) \ket{\vK'} \equiv \partial_{\theta'} \phi_{\vK'}.
\ee

As a crucial observation from \eqref{eq:diff}, at weak fields with $k_F \gg |\ve{M}_{x,y}|\sim \sqrt{B}$, the momentum $\vK'(\theta,\theta')$ rapidly oscillates. Even small variations in $(\theta,\theta')$ causes $\vK'$ to sweep the entire mBZ. As such, we can replace $\partial_{\theta'}\phi_{\vK'(\theta,\theta')}$ with its local average over $\theta$ and $\theta'$, given by $\sum_{\ve K'} \partial_{\theta'}\phi_{\vK'}/N_{\Phi}$. With this substitution,  we arrive at
\be
S_{\rm LP} = -\frac{\omega_c}{2N_{\Phi}}  \sum_{i,j=1}^{N_{\Phi}}\int dt  \int \frac{d\theta d\theta'}{4\pi^2} \,\partial_\theta \phi_{i}\, \partial_{\theta'} \phi_{j}\,  F(\theta-\theta'),
\label{eq:LP}
\ee
where we have used shorthands $i$ and $j$ for the summation over $\vK$ and $\vK'$.
Therefore, in a weak field, $S_{\rm LP}$ corresponds to all-to-all coupling in the mBZ.

Combining Eq.~\eqref{eq:LP} with the free-fermion action in Eq.~\eqref{eq:free}, we obtain the bosonized action for a 2d Fermi liquid in a weak magnetic field, $S= S_{\rm cb} + S_{\rm w} + S_{\rm LP}$.  In the following sections, we will focus on the quantization of the bosonzied theory.

\section{Kohn's theorem and cyclotron resonance}
\label{sec:kohn}

Without loss of generality, in this work we assume the system is invariant under Galilean transformations, which constrains the kinetic energy to be $\epsilon(\vk) = \vk^2/(2m_0)$, where $m_0$ is the bare mass of the electron.  In a Fermi liquid theory, it is well-known~\cite{baym2008landau,DDMS2022} that the effective mass $m$ and $m_0$ are related through the Landau parameter $F_1$:
\be
\frac{1}{m_0} = \frac{1+F_1}{m}\textrm{, where }F_1 = \int\frac{d\theta}{2\pi} F(\theta,0) \cos\theta.
\label{eq:m0}
\ee

In the presence of a magnetic field, Kohn's theorem~\cite{Kohn1959} states that in a Galilean invariant system, there exists a cyclotron mode corresponding to the common motion of the entire system, whose frequency is unrenormalized by interactions. In the quantized bosonic theory, the corresponding statement is that there exists a bosonic excitation in the symmetric sector of all $\phi_{\vK}$ modes, whose energy is 
\be
\omega_c^{(0)} = \frac{B}{m_0}.
\ee

To verify that our bosonized theory is consistent with Kohn's theorem, we  perform a mode expansion for $\phi_i$
\be
\phi_i(\theta) = q_i+ p_i\theta +\sum_{n\neq 0}\frac{a_{i,n}}{\sqrt{|n|}}e^{{i} n\theta},~\theta\in[-\pi,\pi).
\label{eq:mode}
\ee
Note that $\phi_i$ is a compact angular variable,  the zero modes $\{q_i\simeq q_i + 2\pi, p_i\}$ are allowed. These zero modes will be analyzed in detail in Sec.~\ref{sec:dhva}.

In this section, we focus on the oscillatory modes $a_{i,n}$. For these modes, $S_{\rm w}=0$~\cite{ye_wang_first}, and we only need $S_{\rm cb} + S_{\rm LP}$ in the bosonized action:
\be
S_{\rm cb} + S_{\rm w}=\sum_{i=1}^{N_{\Phi}} \int \frac{dt d\theta} {4\pi} \partial_\theta \phi_i \(\partial_t \phi_i - \omega_c \partial_\theta \phi_i\)  - \frac{\omega_c }{2N_{\Phi}} \sum_{i,j=1}^{N_{\Phi}} \int \frac{d\theta d\theta'}{4\pi^2} F(\theta,\theta')\partial_\theta \phi_i \partial_{\theta'}\phi_j.
\ee 
Canonical quantization of the theory leads to the following commutation relation
\be
[a_{i,n}, a_{j,m}] = \delta_{ij} \delta_{n,-m}.
\ee

Due to the rotational invariance, $F(\theta,\theta')$ depends only on $\theta-\theta'$, and thus  $a_{i,n}$  only couple to  $a_{j,-n}$ in the action. Let's focus on the modes with $n=1$. In a Fermi gas, these modes correspond to the energy of an electronic transition between two neighboring Landau levels~\cite{ye_wang_first}. From the Hamiltonian (using the fact that $a_{i,1}^\dagger = a_{i,-1}$), we get
\begin{align}
H_{n=1} 
=& \sum_{i} \omega_c a_{i,1}^\dagger a_{i,1} + \frac{\omega_c}{N_{\Phi}} \sum_{ij} a_{i,1}^\dagger a_{j,1}  \int \frac{d\theta d\theta'}{4\pi^2} F(\theta-\theta') \cos(\theta' - \theta),
\end{align}
where we have used the fact that $F(\theta-\theta')$ is an even function. Let's focus on the symmetric eigenmode corresponding to the synchronous collective motion of $N_{\Phi}$ species
\be
a_{1,\rm sym} = \frac{1}{\sqrt{N_{\Phi}}}\sum_{i=1}^{N_{\Phi}} a_{i,1},
\ee
and the corresponding Hamiltonian is
\begin{align}
H_{1,\rm sym} =& {\omega_c} \(1+\int \frac{d\theta d\theta'}{4\pi^2} F(\theta-\theta') \cos(\theta' - \theta) \) a_{1,\rm sym}^\dag a_{1,\rm sym} \non \\
=& {\omega_c} \(1+F_1 \) a_{1,\rm sym}^\dag a_{1,\rm sym} =  {\omega_c^{(0)}} a_{1,\rm sym}^\dag a_{1,\rm sym}.
\end{align}
where in the last step we used Eq.~\eqref{eq:m0}. We see that indeed the energy of a single quantum of $a_{1,\rm sym}^\dag$ is given by
\be
\Omega_1 = \omega_c (1+F_1) = \omega_c^{(0)},
\ee
consistent with Kohn's theorem.

We can further obtain the energy of higher-order cyclotron resonance, given by modes with $n>1$. For free fermions, these modes represent the fermionic transition between $m$-th and $(m+n)$-th Landau levels. In the Hamiltonian, the corresponding terms are
\begin{align}
H_{n}
=& \sum_{i} n \omega_c a_{i,n}^\dagger a_{i,n} + \frac{n\omega_c}{N_{\Phi}} \sum_{ij} a_{i,n}^\dagger a_{j,n}  \int \frac{d\theta d\theta'}{4\pi^2} F(\theta-\theta') \cos\[n(\theta' - \theta)\].
\end{align}
Focusing on the symmetric eigenmode, we get
\begin{align}
H_{n,\rm sym} =& {n\omega_c} \(1+\int \frac{d\theta d\theta'}{4\pi^2} F(\theta-\theta') \cos\[n(\theta' - \theta)\] \)a_{1,\rm sym}^\dag a_{1,\rm sym} \non \\
=& n{\omega_c} \(1+F_n \) a_{1,\rm sym}^\dag a_{1,\rm sym},
\end{align}
where the Landau parameter $F_n$ is defined as
\be
F_n \equiv  \int \frac{d\theta }{2\pi} F(\theta) \cos(n\theta).
\ee
Together with Kohn's theorem, we obtain that the energy of the higher-order cyclotron resonance is
\be
\Omega_n\equiv n\,\omega_c (1+F_n) = n\,\omega_c^{(0)}\frac{1+F_n}{1+F_1}.
\label{eq:omegan}
\ee 
For $n=1$, we recover the Kohn's theorem.
A similar result was first discussed in 3d Fermi liquid in Ref.~\cite{mermin-cheng-1968}. 

Note that in our present analysis, we retain only terms up to quadratic order in $\phi$. Beyond this, non-Gaussian terms in $S_{\rm LP}$ arise from nonlinear corrections to Eq.~\eqref{eq:df}, leading to damping of the cyclotron resonance. The cyclotron resonance modes can be interpreted as angular harmonics of Fermi surface deformation. In the absence of a magnetic field, their damping rates have been predicted to show a dichotomy between even and odd harmonics~\cite{levitov-2019}. It would be interesting to examine whether this effect persists in the presence of a magnetic field~\cite{2024arXiv240905147M}.

\section{Specific heat}
\label{sec:ct}
As we mentioned, the computation of the specific heat from bosonization of a 2d FS is challenging due to the issue of UV/IR mixing~\cite{DDMS2022}.
In the presence of a magnetic field, $\ell_B$ makes the theory UV complete, and the specific heat of a 2d Fermi gas can be obtained rigorously from the bosonic approach~\cite{ye_wang_first}. In this treatment, the thermal energy of a Fermi gas predominantly comes from to that of the black body radiation in the oscillator sector $\{a_{i,n}\}$. Here we apply the bosonic approach a 2d Fermi liquid.

In a Fermi liquid, the Landau-parameter term couples the $N_{\Phi}$ species of bosons. Similar to the $n=1$ modes we considered in Sec.~\ref{sec:kohn}, in general the Hamiltonian for the $a_{i,n}$ oscillatory modes is
\begin{align}
H_{n}
=& \sum_{i} n \omega_c a_{i,n}^\dagger a_{i,n} + \frac{n\omega_c}{N_{\Phi}} \sum_{ij} a_{i,n}^\dagger a_{j,n}  \int \frac{d\theta d\theta'}{4\pi^2} F(\theta-\theta') \cos(\theta' - \theta)\nonumber\\
=& n\omega_c \sum_{ij} a_{i,n}^\dag \mathbb{M}^{(n)}_{ij} a_{j,n},
\end{align}
where $\mathbb{M}^{(n)}$ is a $N_{\Phi}\times N_{\Phi}$ matrix given by
\be
\mathbb{M}^{(n)} = \mathbb{I} + \frac{F_n}{N_{\Phi}} \(\begin{array}{cccc}
     1& 1&\cdots \\
     1& 1& \\
     \vdots &&\ddots 
\end{array}\)_{N_{\Phi}}\equiv \mathbb{I} + \frac{F_n}{N_{\Phi}}\mathbb{J},
\label{eq:Mn}
\ee
where $\mathbb{I}$ is the $N_{\Phi}\times N_{\Phi}$ identity matrix and  $\mathbb{J}$ is a $N_{\Phi}\times N_{\Phi}$ ``all-one" matrix. $\mathbb{J}$ is a highly singular matrix, with one eigenvalue being $N_{\Phi}$ and the rest being all zero. Therefore, after diagonalizing $\mathbb{M}^{(n)}$ and thus $H_{n}$, the thermal energy at the temperature $T=1/\beta$ is given by the Bose-Einsten distribution:
\be
E_n = (N_{\Phi} -1)\frac{n\omega_c}{e^{\beta n\omega_c}-1} + \frac{n\omega_c (1+F_n)}{e^{\beta n\omega_c(1+F_n)}-1},
\label{eq:EnFn}
\ee
where the second term corresponds to the $n$-th cyclotron resonance mode in Eq.~\eqref{eq:omegan}.
Taking the thermodynamic limit $L\to \infty$, the thermal energy is dominated by the first term of \eqref{eq:EnFn}:
\be
E_n \approx N_{\Phi}\frac{n\omega_c}{e^{\beta n\omega_c}-1}.
\ee


The total thermal energy is given by
\be
E_T\approx  N_{\Phi} \sum_{n=1}^{\infty} \frac{n\omega_c}{e^{\beta n\omega_c}-1} \approx N_{\Phi} \int_{0}^{\infty} \frac{dx x \omega_c}{e^{x\beta \omega_c} -1},
\ee
which is nothing but the thermal energy of black body radiation in 1d. In the second step, we have taken the limit $\omega_c\ll T$, and replaced the summation with an integral. 
  Evaluating the integral, we get for the thermal energy and the specific heat that is independent of the magnetic field
\be
E_T = \frac{\pi^2}{6}g(\epsilon_F) T^2,~~~C(T) = \frac{\pi^2}{3}g(\epsilon_F)T,
\label{eq:ct}
\ee
where $g(\epsilon_F) = mL^2/2\pi$ is the density of states at Fermi energy. This is a key result of the Fermi liquid theory --- the specific heat of a Fermi liquid has the same form as that of a 2d Fermi gas, with $m$ being the effective mass.

\section{De Haas-van Alphen effect}
\label{sec:dhva}
It has been recently shown~\cite{ye_wang_first} that in a 2d Fermi gas, the dHvA effect, i.e., the nonanalytic oscillatory behavior of the free energy, is completely captured in the bosonic theory by the zero-mode sector $\{q_i,p_i\}$ of the mode expansion in Eq.~\eqref{eq:mode}. In this section we develop a bosonic theory of dHvA in a 2d Fermi liquid.

We focus on the action obtained by taking $\phi_i = q_i + p_i\theta$ in Eqs.~(\ref{eq:free}, \ref{eq:LP}):
\be
S_{\rm zero}[p,q] = \int dt \[\sum_{i=1}^{N_{\Phi}}\(-\frac{\mathcal{A}_{\rm FS}}{2\pi B}\dot q_i+p_i\dot q_i-\frac{ \omega_c}{2}p_i^2\)-\frac{\omega_c F_0}{2N_{\Phi}}\sum_{ij}^{N_{\Phi}} p_i p_j\],
\ee
where $F_0 = \int F(\theta)\, d\theta /(2\pi)$ is the zeroth Landau parameter. We see that the area enclosed by the FS naturally enters via integrating over the single-particle distribution function
\be
\mathcal{A}_{\rm FS} = \int d\vk\, f_0(\vk).
\label{eq:single}
\ee
We note that the term $-{\mathcal{A}_{\rm FS}\dot q}/(2\pi B)$ term in $S_{\rm zero}$ is a topological $\theta$-term~\cite{altland}, which does not enter the equation of motion, but affect the energy spectrum of the system by altering the quantization of $\hat p_i$. In this regard, the dHvA effect is emblematic of a topological phenomenon. The quantization condition can either be obtained directly from performing the path integral in imaginary time~\cite{ye_wang_first}, or by analogy with the Aharonov-Bohm effect for a charged particle moving on a ring in the presence of a magnetic flux~\cite{altland}. As a result of the $\theta$-term, the spectrum of $\hat p_i$ is 
\be
\operatorname{spec}\({\hat p}_i\) = \mathbb{Z} -  \frac{\mathcal{A}_{\rm FS}}{2\pi B}.
\ee

The corresponding Hamiltonian is therefore
\be
\hat H_{\rm zero} = \frac{\omega_c}{2}\sum_{i=1}^{N_{\Phi}} \(\tilde p_i+\frac{\mathcal{A}_{\rm FS}}{2\pi B}\)^2 + \frac{\omega_c F_0}{2N_{\Phi}}\sum_{ij}^{N_{\Phi}} \(\tilde p_i+\frac{\mathcal{A}_{\rm FS}}{2\pi B}\)\(\tilde p_j+\frac{\mathcal{A}_{\rm FS}}{ 2\pi B}\)
\label{eq:27}
\ee
where $\tilde p_i\in\mathbb{Z}$. As long as $1+F_0>0$, this Hamiltonian is semi-positive-definite, which coincides with a stability criterion of the Fermi liquid~\cite{baym2008landau}. 

Physically, $\hat H_{\rm zero}$ counts the energy cost of introducing extra quasiparticles/holes to the system in the grand canonical ensemble. In the thermodynamic limit $N_{\Phi}\to \infty$, the second term in Eq.~\eqref{eq:27} can be neglected for $p_i=\mathcal{O}(1)$, and the excitation energy of $\hat H_{\rm zero}$ is given by
\be
\Delta E_{\rm zero}(\{p_i\}) = \frac{\omega_c}{2}\sum_{i=1}^{N_{\Phi}} \[p_i^2 + \frac{p_i}{\pi} \Delta\(\frac{\mathcal{A}_{\rm FS}}{B}\)\]
\label{eq:qps}
\ee
where $p_{i} \in \mathbb{Z}$, and $\Delta(x)$ is $x$ modulo $2\pi$ taken in $[-\pi,\pi)$.\footnote{For compactness, we have replaced $\tilde p_i$ with  $p_i$. It is related but not to be confused with the dynamical zero mode $p_i$ in Eq.~\eqref{eq:mode}.} It can be directly verified that $\Delta E_{\rm zero}(\{p_i\})$ precisely matches the total energy of $|p_i|$ quasiparticles/holes (depending on the sign of $p_i$) in each magnetic momentum channel $i$; specifically, the first term represents the total energy when the chemical potential lies right in between two Landau levels, and the second term corrects for the deviation of the chemical potential from that midpoint. 

It is clear that $\hat H_{\rm zero} $ and $\Delta E_{\rm zero}$ both exhibit  oscillatory behavior periodic in $\mathcal{A}_{\rm FS}/B$, which is the origin of the dHvA effect.
Strictly speaking, the bosonization procedure does not fix the ground state energy of the system, which is not measurable within the path integral, and does not exclude an extra additive c-number to $\hat H_{\rm zero}$ that is by itself oscillatory in $\mathcal{A}_{\rm FS}/B$. However, from the fermionic side we know that dHvA oscillations becomes exponentially small for $T\gtrsim \omega_c$, which, as we shall see, can only be true if there are no additional oscillatory terms in the ground state energy.

We note that $H_{\rm zero}$ also contributes to the specific heat. In particular, the oscillatory behavior we discuss below also causes oscillations in specific heat as a function of $B$. However, this contribution is negligible for $\omega_c\ll T$ compared with Eq.~\eqref{eq:ct}.

\subsection{Warmup: dHvA for 2d Fermi gas from bosonization}
\label{sec:warmup}

It is instructive to begin with dHvA of a 2d Fermi gas~\cite{ye_wang_first,ye_wang_second} with $F_0=0$ in Eq.~\eqref{eq:27}. As we discussed, $p_i\in \mathbb{Z}$ corresponds to varying the number of fermions, and thus they are summed over in the grand partition function.  We write down the part of the grand potential containing dHvA oscillations as 
\begin{align}
\Omega_{\rm osc} =& -T \log \left\{ \sum_{\{p_i\in \mathbb{Z}\}}\exp\[-\frac{\beta\omega_c}{2}\sum_{i=1}^{N_{\Phi}}\(p_i+\frac{\Delta(\mathcal{A}_{\rm FS}/B)}{2\pi}\)^2\]\right\}.
\end{align}
 In the absence of interactions, the summation in the argument of the log factorizes:
\be
\Omega_{\rm osc}= -T \log \left\{ \sum_{p\in\mathbb{Z}}\exp\[-\frac{\beta\omega_c}{2}\(p+\frac{\Delta(\mathcal{A}_{\rm FS}/B)}{2\pi}\)^2\]\right\}^{N_{\Phi}}.
\label{eq:29new}
\ee

After applying the Possion resummation formula $\sum_{p\in \mathbb{Z}} f(p) = \sum_{k\in\mathbb Z} \int f(x)\,e^{2\pi i k x}\,dx$, we get
\begin{align}
\Omega_{\rm osc}=& -N_{\Phi}T \log \left\{\sum_{n\in\mathbb{Z}} e^{-i n\Delta}\exp\(-\frac{2\pi^2 k^2 T }{\omega_c}\)\right\}\nonumber\\
=&  -N_{\Phi}T \log \[\theta_3\(\frac{\Delta}2,\kappa\)\],
\end{align}
where 
\be
\Delta = \Delta\(\frac{\mathcal{A}_{\rm FS}}{B}\) = \frac{\mathcal{A}_{\rm FS}}{B} \bmod 2\pi \in [-\pi,\pi),~~~\kappa=\exp\(-\frac{2\pi^2 T}{\omega_c}\),
\label{eq:q}
\ee 
and we have used the definition of the Jacobi $\theta$-function~\cite{weisstein2000jacobi}. It is well-known that the Jacobi function $\theta_3$ has a triple-product representation:
\beq
\theta_3\(\frac{\Delta}{2},\kappa\)=\prod_{m=1}^\infty(1-\kappa^{2m})\(1+\kappa^{2m-1}e^{i\Delta}\)\(1+\kappa^{2m-1}e^{-i\Delta}\),
\eeq
which can be used to reexpress the grand potential as
\be
\Omega_{\rm osc} = -N_{\rm \Phi} T \Re{\sum_{m=1}^{\infty} \log\(1+\kappa^{2m-1}e^{i\Delta}\)} + \cdots,
\ee
where $\cdots$ stands for nonoscillatory terms and will be dropped hereafter. Further expanding the log, we get
\be
\Omega_{\rm osc} = N_{\rm \Phi} T \Re{\sum_{k=1}^{\infty} \sum_{m=1}^{\infty} \frac{(-)^k \kappa^{(2m-1 )k} e^{i k\Delta}}{k}}.
\ee
Summing over $m$, plugging in the definition of $\kappa$ in Eq.~\eqref{eq:q}, and taking the real part, we get
\be
\Omega_{\rm osc} = N_{\Phi} T \sum_{k=1}^{\infty} \frac{(-)^k \cos(k\mathcal{A}_{\rm FS}/B)}{k\sinh(2\pi^2 k T/\omega_c)} \equiv \sum_{k=1}^{\infty} A_k \cos\(\frac{k\mathcal{A}_{\rm FS}}B\),
\label{eq:LK}
\ee
which is the celebrated LK formula~\cite{LK1956}. We show the first two amplitudes $A_{1,2}$ as a function of temperature $T$ in Fig.~\ref{fig:1}. For $T\gg\omega_c$, the oscillations vanish as expected, confirming that the Hamiltonian in Eq.~\eqref{eq:27} without any additional c-number terms fully captures the dHvA oscillations.

It is also instructive to evaluate $\Omega_{\rm osc}$ separately in two limiting cases with $T\ll \omega_c$ and $T\gtrsim \omega_c$, which we do below. 

\subsubsection{$T\ll \omega_c$} \label{sec:freelow}


At $T=0$, the free energy is simply the ground state energy, obtained by setting $p=0$ in Eq.~\eqref{eq:29new}:
\be
\Omega_{{\rm osc}}(T=0)= \frac{N_{\Phi}\omega_c}{4\pi}  \[\Delta\(\frac{\mathcal{A}_{\rm FS}}{ B}\)\]^2,
\label{eq:26}
\ee
which exhibits dHvA oscillations periodic in $1/B$ with a period of $2\pi / \mathcal{A}_{\rm FS}$. Rewriting the oscillatory part as a Fourier series in  $\cos(k\Delta)$, we get
\be
\Omega_{\rm osc}(T=0) = \sum_{k=1}^\infty A_k(T=0) \cos(k\Delta)\textrm{, where } A_k(T=0)= N_{\Phi}\omega_c \,\frac{(-)^k}{2\pi^2 k^2},
\label{eq:t0}
\ee
in agreement with the LK formula \eqref{eq:LK}.

For low temperatures with $T\ll \omega_c$, it suffices to only include the ground state and the lowest excited state for each bosonic species, we get from Eq.~\eqref{eq:29new} by setting $p=-\sgn(\Delta)$:
\begin{align}
\Omega_{\rm osc}(T) &\approx \Omega_{\rm osc}(T=0) - N_{\Phi}T \log \(1+\exp\left\{-\frac{\beta\omega_c}{2}\[\(1-\frac{|\Delta|}{2\pi}\)^2-\(\frac{|\Delta|}{2\pi}\)^2\]\right\}\)\non\\
&\equiv \Omega_{\rm osc}(T=0) + \Omega_{\rm th}(T).
\label{eq:oT0}
\end{align}
$\Omega_{\rm th}$ can then be expanded as
\be
\Omega_{\rm th}= N_{\Phi}T\sum_{m=1}^\infty \frac{(-)^m}m \exp\[-\frac{\beta m\omega_c}{2}\(1-\frac{|\Delta|}{\pi}\)\]
\label{eq:oT}
\ee
For a general value of $|\Delta|$, $\Omega_{\rm th}$ is exponentially suppressed. However, at special values of $\Delta = \pm \pi$ the exponential factor is eliminated. As we shall see, this leads to a power-law dependence of $\Omega_{\rm th}$ on $T$. On the other hand, for any higher excited states, their contribution to $\Omega_{\rm th}$ is always exponentially suppressed by $\exp(-\alpha \omega_c/T)$ whose derivatives with respect to $T$ vanishes to all orders,  and  is thus negligible for $T\ll \omega_c$.

We can express $\Omega_{\rm th}$ as a cosine series with basis functions $\cos(k\Delta)$, where we remind $\Delta \in [-\pi,\pi)$. The $k$-th Fourier component is
\begin{align}
A_{{\rm th},k} = \frac{2N_{\Phi}T}\pi \sum_{m=1}^\infty\frac{(-)^m}{m}\exp(-\frac{\beta m\omega_c }2)\underbrace{\int_0^{\pi}\exp\(\frac{\beta m\omega_c|\Delta|}{2\pi}\)\cos(k\Delta) d\Delta}_{I_m}.
\label{eq:oTk}
\end{align}
For $\beta\omega_c\gg 1$, the dominant contribution to $I_m$ comes from $\Delta \approx  \pi$.
Up to exponetially small corrections,  we can extend the lower cutoff in $I_m$ to $-\infty$, and
\be
I_m = \Re \frac{(-)^n\exp(\frac{\beta\omega_c m}2)}{\frac{\beta\omega_c m}{2\pi}+ik} =\frac{(-)^n 2\pi\beta \omega_c m\exp(\frac{\beta\omega_c m}2)  }{\beta^2\omega_c^2m^2+4\pi^2k^2}.
\ee
Inserting $I_m $ back into Eq.~\eqref{eq:oTk}, we see that the exponential suppression in $A_{{\rm th},k}$ is compensated. 
We get
\begin{align}
A_{{\rm th},k} =& 4N_{\Phi} \omega_c (-)^k \sum_{m=1}^\infty\frac{(-)^m}{\beta^2\omega_c^2m^2+4\pi^2k^2} \non\\
=& 4N_{\Phi} \omega_c (-)^k \[\frac{T}{4n\omega_c\sinh(2\pi^2kT/\omega_c)}-\frac{1}{8\pi^2 k^2}\] \non\\
=& - \frac{(-)^{k}N_{\Phi}\pi^2T^2}{3\omega_c} + \mathcal O(T^4).
\label{eq:atten}
\end{align}

The  temperature-dependent $A_k(T\ll \omega_c)$ is given by $A_k(T=0)+A_{{\rm th},k}(T)$. The last line of Eq.~\eqref{eq:atten} captures the attenuation behavior of the dHvA at low tempeartures, consistent with the LK formula \eqref{eq:LK}.\footnote{We note that, from the second line in Eq.~\eqref{eq:atten}, the formula actually reproduces the full LK formula for \emph{all} temperatures (that are small compared to the Fermi energy). This is accidental ---   in the procedure above we have dropped  $\sim \exp(-\alpha \omega_c/T)$ terms with vanishing derivatives in $T$, and  the full LK formula \eqref{eq:LK} happens to not contain any of these terms.}

\subsubsection{$T\gtrsim \omega_c$} \label{sec:freehigh}

Instead of factorizing the summation, let us rewrite $\Omega_{\rm osc}$ using the Poisson resummation formula as
\be
\Omega_{\rm osc} = -T \log\[\sum_{\{k_i\in \mathbb Z\}}\exp(- i \Delta \sum_{i=1}^{N_{\Phi}}k_i) \exp(-\frac{2\pi^2 T}{\omega_c}\sum_{i=1}^{N_{\Phi}}k_{i}^2)\] ,
\label{eq:29free}
\ee

When $T\gtrsim \omega_c$, we can expand the argument of the log as 
\begin{align}
\Omega_{\rm osc}  = -T \log&\[1+2N_{\Phi}\kappa\cos(\Delta) +2\frac{N_{\Phi}(N_{\Phi}-1)}{2}\kappa^2\cos(2\Delta) \right.
\nonumber\\
&\left.+ 2\frac{N_{\Phi}(N_{\Phi}-1)(N_{\Phi}-2)}{6}\kappa^3\cos(3\Delta)+\cdots\],
\label{eq:64}
\end{align}
where \be
\kappa=\exp(-\frac{2\pi^2 T}{\omega_c})\ll 1,
\ee
and for each harmonic  ($\cos k\Delta$) we have only kept the term lowest order in $\kappa$, which is exponentially small. Within the logarithm, the first term of Eq.~\eqref{eq:64} arises when all $k_i=0$, the second term when exactly one $k_i$ takes the value $\pm 1$ (the rest zero), the third term when either two $k_i$ take $+1$ or two take $-1$, and the final term when either three $k_i$ take $+1$ or $-1$.

Although $\Omega_{\rm osc}$ is an extensive quantity, we see that the terms in the argument of the log are not in general so. Upon Taylor expanding the logarithm, terms with higher powers of $N_{\Phi}$ must cancel. 
Thus, as a shortcut, we need to only keep track of the $\mathcal{O}(N_{\Phi})$ terms, which in this situation are only contained in the linear order in the Taylor expansion $\log (1+y) =y+\cdots$. This leads to 
\be
\Omega_{\rm osc} =\sum_{k=1}^\infty A_k\cos(\frac{k\mathcal{A}_{\rm FS}}{B})\textrm{, where } A_k=2N_{\Phi}T\frac{(-)^k}{k}\exp(-\frac{2\pi^2k T}{\omega_c}) .
\label{eq:supp}
\ee
 Eq.~\eqref{eq:supp} gives the correct high-temperature asymptotic behavior of the LK formula \eqref{eq:LK}. As we shall see, by similarly leveraging the extensiveness of $\Omega_{\rm osc}$ we can obtain the high-temperature behavior of dHvA for a 2d Fermi liquid.

\subsection{dHvA for a 2d Fermi liquid: $T\ll \omega_c$}

We now turn to the case of a 2d Fermi liquid with Landau parameters.
For $T=0$, dHvA  comes from to the oscillation of ground state energy of $H_{\rm zero}$.  It is straightforward to see from Eq.~\eqref{eq:27} that the ground state is obtained by setting all $\tilde p_i=0$, and the oscillatory part is 
\be
\Omega_{{\rm osc}}(T=0)= \frac{N_{\Phi}\omega_c}{4\pi} (1+F_0) \[\Delta\(\frac{\mathcal{A}_{\rm FS}}{ B}\)\]^2,
\label{eq:26new}
\ee
where we remind $\Delta(x)$ is $x$ modulo $2\pi$ taken in $[-\pi,\pi)$. This result is simply the dHvA for a $T=0$ Fermi gas rescaled by $1+F_0$. From Eq.~\eqref{eq:t0}, the Fourier component is given by
\be
A_k(T=0)= N_{\Phi}\omega_c(1+F_0) \,\frac{(-)^k}{2\pi^2 k^2}.
\label{eq:AkT0}
\ee

We can further compute the attenuation from finite temperature effects. 
The grand potential of the zero modes is given by
\begin{align}
\Omega_{\rm osc} =& -T \log \left\{ \sum_{\{p_i\in \mathbb{Z}\}}\exp\[-\frac{\beta\omega_c}{2}\sum_{i=1}^{N_{\Phi}}\(p_i+\frac{\Delta(\mathcal{A}_{\rm FS}/B)}{2\pi}\)^2\] \right.\nonumber\\
&~~~~~~~~~~~~~\left.\times\exp\[-\frac{\beta\omega_cF_0}{2 N_{\Phi}}\sum_{i,j=1}^{N_{\Phi}}\(p_i+\frac{\Delta(\mathcal{A}_{\rm FS}/B)}{2\pi}\)\(p_j+\frac{\Delta(\mathcal{A}_{\rm FS}/B)}{2\pi}\)\]\right\}.
\label{eq:gp}
\end{align}

Just like the free-fermion case in Eq.~\eqref{eq:oT0}, the leading  contribution to finite-$T$ attenuation of $A_k$ for $T\ll \omega_c$ can be obtained by restricting $p_i\in \{0,-\sgn(\Delta)\}$, where we remind $\Delta$ is a shorthand for $\Delta(\mathcal{A}_{\rm FS}/B)$. However, here the summation over $\{p_i\}$ does not factorize. We can instead express the finite-$T$ correction as
\begin{align}
\Omega_{\rm th} \approx & -T \log\left\{\sum_{m=0}^{N_\phi}\binom{N_\phi}{m}\exp[-\frac{\beta\omega_c}{2}\(1-\frac{|\Delta|}{\pi}\)m]\right. \non\\
&~~~~~~~~~~~~~\left.\times \exp[-\frac{\beta\omega_cF_0}{2}\(1-\frac{|\Delta|}{\pi}\)^2\frac{m^2}{N_{\Phi}} -\frac{\beta\omega_cF_0}{2}\(\frac{|\Delta|}{\pi}-\frac{|\Delta|^2}{\pi^2}\)m ]\right\},
\label{eq:72}
\end{align}
where $m$ counts the number of $p_i$'s being set to nonzero. For $F_0=0$, using the binomial theorem, we recover Eq.~\eqref{eq:oT0}.

For a nonzero $F_0$, there is no closed-form expression of $\Omega_{\rm th}$ for $T\neq 0$ in terms of elementary functions. However, for $T\ll \omega_c$, we observe that, just like the free fermion case, the dominant contribution to the Fourier transformation of $\Omega_{\rm th}$ comes from $1-|\Delta|/\pi \sim T/\omega_c \ll 1$. Therefore, up to  corrections that are higher-order in $T$, the Fourier coefficient $A_{{\rm th},k}$ of $\Omega_{\rm th}$ is identical to that of 
\begin{align}
\Omega_{\rm th}' &= -T \log\left\{\sum_{m=0}^{N_\phi}\binom{N_\phi}{m}\exp[-\frac{\beta m\omega_c}{2}\(1-\frac{|\Delta|}{\pi}\)(1+F_0)]\right\}\non \\ 
 &= - N_{\Phi}T \log \left\{1+\exp\[-\frac{\beta\omega_c}{2}\(1-\frac{|\Delta|}{\pi}\)(1+F_0)\]\right \},
\label{eq:72'}
\end{align}
with terms of higher order in $1-|\Delta|/\pi$ dropped in the exponent. 
Following the same steps as in Sec.~\ref{sec:freelow}, we get
\begin{align}
A_{{\rm th},k}=& - \frac{(-)^{k}N_{\Phi}\pi^2T^2}{3\omega_c (1+F_0)} + \mathcal O(T^3).
\label{eq:zeta}
\end{align}

The  temperature-dependent $A_k(T\ll \omega_c)$ is given by $A_k(T=0)+A_{{\rm th},k}$. From Eqs.~(\ref{eq:AkT0}, \ref{eq:zeta}), we see that a positive Landau parameter $F_0$ not only enhances the amplitude of dHvA for all harmonics, but also weakens their finite-$T$ attenuation behavior. Furthermore, compared to the free fermion case, $A_k(T\ll \omega_c)$ for a Fermi liquid is obtained from the free fermion result by replacing $\omega_c$  by 
\be
\omega_{c,0} = \omega_c (1+F_0) = \omega_c^{(0)} \frac{1+F_0}{1+F_1}.
\ee
in Eq.~\eqref{eq:LK}. This result closely parallels the expression for the cyclotron resonance in Eq.~\eqref{eq:omegan}, with the zero modes $\{p_i,q_i\}$ viewed as the $n=0$ sector.

To the best of our knowledge, no result for $A_k(T \ll \omega_c)$ is known from the fermionic approach~\cite{maslov-2003}, due to difficulties caused by the oscillatory behavior in the fermionic self energy.

\subsection{dHvA for a 2d Fermi liquid: $T\gtrsim \omega_c$ }
For $T \gtrsim \omega_c$, thermal smearing leads to exponential suppression of the oscillations. To see this, we apply the Poisson resummation formula to Eq.~\eqref{eq:gp} and obtain the oscillatory part as
\be
\Omega_{\rm osc} = -T \log\[\sum_{\{k_i\in \mathbb Z\}}\exp(- i \Delta \sum_{i=1}^{N_{\Phi}}k_i) \exp(-\frac{2\pi^2 T}{\omega_c}\sum_{i,j=1}^{N_{\Phi}}k_{i}(\mathbb{M}^{(0)})^{-1}_{ij}k_j)\] ,
\label{eq:29}
\ee
where (cf.~Eq.~\eqref{eq:Mn})
\be
\mathbb{M}^{(0)}=\mathbb{I} + \frac{F_0}{N_{\Phi}}\mathbb{J},~~~ \(\mathbb{M}^{(0)}\)^{-1}=\mathbb{I} -\frac{F_0}{1+F_0}\frac{\mathbb{J}}{N_{\Phi}},
\ee
and we remind that $\mathbb{J}$ is a $N_{\Phi} \times N_{\Phi}$ all-one matrix.\footnote{Interestingly, the argument of the log in Eq.~\eqref{eq:29} can be reexpressed as the  Riemann theta function $\Theta(\mathfrak{D},\mathfrak{N})$~\cite{weisstein2000riemann}, which is a generalization of the Jacobi theta function to matrix variables, with $\mathfrak{D} = -\Delta (1,1,\cdots)/(2\pi)$ and $\mathfrak{N}=iN{2\pi T}/{\omega_c}$. However, we are  not aware of any properties of this special function that are useful for our purposes; unlike the Jacobi theta function, $\Theta(\mathfrak{D},\mathfrak{N})$ does not have a product form. This has prevented us from obtaining a closed-form solution for dHvA in a 2d Fermi liquid applicable to all temperatures.}

We can leverage the extensiveness\footnote{Although we do not directly prove the extensiveness of $\Omega_{\rm osc}$, via a straightforward perturbative expansion of Eq.~\eqref{eq:74}, one can verify that nonextensive terms such as $\sim N_{\Phi}^2\kappa^3 \chi$ indeed cancel.} of $\Omega_{\rm osc}$ to obtain the dHvA amplitudes at $T\gtrsim \omega_c$.
Similar to Eq.~\eqref{eq:64}, the expansion inside the logarithm becomes
\begin{align}
\Omega_{\rm osc}  = -T \log&\[1+ N_{\Phi}(\bar\kappa+ \bar \kappa^*)\exp(\frac{\chi}{N_{\Phi}}) +\frac{N_{\Phi}(N_{\Phi}-1)}{2}(\bar\kappa^2+\bar\kappa^{*2})\exp(\frac{4\chi}{N_{\Phi}}) \right.\non\\
&\left.+ \frac{N_{\Phi}(N_{\Phi}-1)(N_{\Phi}-2)}{6}(\bar\kappa^3+\bar\kappa^{*3})\exp(\frac{9\chi}{N_{\Phi}})+\cdots\],
\label{eq:74}
\end{align}
where we have defined
\be
\bar\kappa = \kappa e^{i\Delta},~~~ \bar\kappa^* = \kappa e^{-i\Delta},
~~~ \chi=\frac{2\pi^2T}{\omega_c}\frac{F_0}{1+F_0}.
\ee 
Inside the logarithm of Eq.~\eqref{eq:74}, the terms have the same origin as those in Eq.~\eqref{eq:64} for free fermions.

Just like in Sec.~\ref{sec:freehigh}, we can again obtain a Taylor expansion for $\Omega_{\rm osc}$ in terms of $\kappa$ and $\bar\kappa$ by only keeping track of the extensive terms. The first harmonic is given by the leading order term in the expansion:
\be
\Omega_1= -N_{\Phi} T(\bar\kappa+ \bar \kappa^*)= -2N_{\Phi} T \exp(-\frac{2\pi^2 T}{\omega_c})\cos(\frac{\mathcal{A}_{\rm FS}}{B}),
\label{eq:A1}
\ee
whose amplitude $A_1$ follows the LK formula \eqref{eq:LK} at $T\gtrsim \omega_c$~\cite{maslov-2003}. 

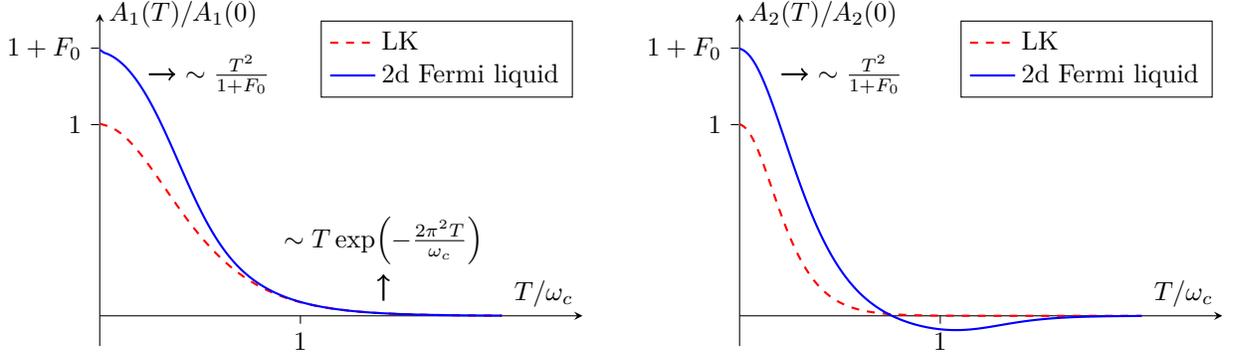
\begin{figure}
  \centering
  \begin{tikzpicture}
    \begin{axis}[
      axis lines=middle,
      every axis x line/.append style={->},
      every axis y line/.append style={->},
      xlabel={$T/\omega_c$},
      ylabel={$A_1(T)/A_1(0)$},
      ylabel style={yshift=3mm},
      domain=0.001:.5,
      xmax=.6,
      ymin = -0.004,
      ymax=0.04,
      samples=100,
      smooth,
      width=8cm,
      height=6cm,
      xtick=\empty,
      xticklabels=\empty,
      ytick=\empty,
      yticklabels=\empty,
      legend style={at={(0.98,0.98)},anchor=north east},
      legend cell align=left,
      after end axis/.code={
        \draw (axis cs:0.25,0) -- ++(0,-3pt) node[below]{1};
         \draw (axis cs:0,0.02533) -- ++(-3pt,0) node[left]{$1$};
         \draw (axis cs:0,0.03546) -- ++(-3pt,0) node[left]{$1+F_0$};
      },
    ]
      \addplot[
        mark=none,
        dashed,
        thick,
        color=red,
      ] { x / (2 * sinh(2 * pi^2 * x)) };
      \addlegendentry{LK}

      \addplot[
        mark=none,
        thick,
        color=blue,
      ] { x / (2 * sinh(2 * pi^2 * x / (1 + 0.4 / ( exp (30*(x-0.1)) + 1)) ) ) };
      \addlegendentry{2d Fermi liquid}

      \draw[->, thick, black]
        (axis cs:0.062,0.032) -- ++(10pt,0pt)
        node[right] {$\sim\frac{T^2}{1+F_0}$};
      \draw[->, thick, black]
        (axis cs:0.353,0.002) -- ++(0pt,10pt)
        node[above] {$\sim T\exp(-\frac{2\pi^2 T}{\omega_c})$};
    \end{axis}
  \end{tikzpicture}
\hspace{5mm}
    \begin{tikzpicture}
    \begin{axis}[
      axis lines=middle,
      every axis x line/.append style={->},
      every axis y line/.append style={->},
      xlabel={$T/\omega_c$},
      ylabel={$A_2(T)/A_{2}(0)$},
      ylabel style={yshift=3mm},
      domain=0.001:.5,
      xmax=.6,
      ymin=-.004,
      ymax=0.04,
      samples=100,
      smooth,
      width=8cm,
      height=6cm,
      xtick=\empty,
      xticklabels=\empty,
      ytick=\empty,
      yticklabels=\empty,
      legend style={at={(0.98,0.98)},anchor=north east},
      legend cell align=left,
      after end axis/.code={
        \draw (axis cs:0.25,0) -- ++(0,-3pt) node[below]{1};
         \draw (axis cs:0,0.02533) -- ++(-3pt,0) node[left]{$1$};
         \draw (axis cs:0,0.03546) -- ++(-3pt,0) node[left]{$1+F_0$};
      },
    ]
      \addplot[
        mark=none,
        dashed,
        thick,
        color=red,
      ] {2 * x / (2 * sinh(4 * pi^2 * x)) };
      \addlegendentry{LK}

      \addplot[
        mark=none,
        thick,
        color=blue,
      ] {2* x / (2 * sinh(4 * pi^2 * x / (1 + 0.4) )  ) * (-14 * tanh(15*(x - 0.3)) -13)};
      \addlegendentry{2d Fermi liquid}

      \draw[->, thick, black]
        (axis cs:0.053,0.032) -- ++(10pt,0pt)
        node[right] {$\sim \frac{T^2}{1+F_0}$};

    \end{axis}
  \end{tikzpicture}
  
  \caption{Schematic plot of the dHvA apmplitudes $A_1(T)$ and $A_2(T)$ of a 2d Fermi liquid, which deviates from the LK formula \eqref{eq:LK} (dashed curves). The quantitative results are given in Eqs.~(\ref{eq:AkT0}, \ref{eq:zeta}, \ref{eq:A1}, \ref{eq:A2}).}
  \label{fig:1}
\end{figure}

The second harmonic $\sim \bar\kappa^2+\bar\kappa^{*2}$ has two contributions: from the second term in the argument of \eqref{eq:74} at quadratic order and from the third term at linear order.\footnote{Note that in the expansion there is no need to keep cross terms $\propto\bar\kappa^n\bar\kappa^{*m}$ with $nm\neq 0$, as they correspond to $\kappa^{m+n}\cos[(m-n)\Delta]$, which is of higher order in $\kappa\ll 1$ for the $(m-n)$-th harmonic.}  Isolating extensive contributions, we get
\begin{align}
\Omega_2=&\frac{N_{\Phi} T}{2} (\bar\kappa^2+\bar\kappa^{*2})(1-2\chi)  =N_{\Phi}T \exp(-\frac{4\pi^2 T}{\omega_c})\(1-\frac{4\pi^2T}{\omega_c}\frac{F_0}{1+F_0}\)\cos(\frac{2\mathcal{A}_{\rm FS}}{B}).
\label{eq:34}
\end{align}
We see that unlike the first harmonic, $A_2$ receives a Fermi liquid correction:
\be
A_2(T\gtrsim \omega_c) = N_{\Phi}T \exp(-\frac{4\pi^2 T}{\omega_c})\(1-\frac{4\pi^2T}{\omega_c}\frac{F_0}{1+F_0}\).
\label{eq:A2}
\ee
This is a key result of this work. Interestingly, due to interaction effects, $A_2$ changes sign at high enough temperatures, which can be directly measured experimentally, from which $F_0$ can be extracted. We summarize our key findings in this section in Fig.~\ref{fig:1}.

From the fermionic perspective, at $T\gtrsim \omega_c$, the oscillatory behavior in the self energy is suppressed, and in Ref.~\cite{maslov-2003}, the amplitude for the leading harmonic $A_1$ was found. Using the ``first-Matsubara rule"~\cite{first-Mats} of the non-oscillatory part of the self energy, $A_1(T)$ is consistent with the LK formula (and our result). The authors pointed out that $A_2(T\gtrsim \omega_c)$ goes beyond the LK formula due to the oscillations in the self energy, but did not obtain its expression within their approach. In comparison, our bosonized approach allows us to obtain, in principle, all $A_k$ at $T\gtrsim \omega_c$ (although the book keeping gets more complicated for larger $k$). Moreover, as we showed, our method also allows for the analysis of dHvA at low temperatures with $T\ll \omega_c$.

\section{Comparison with a 3d Fermi liquid}
\label{sec:3d}

We have shown that the dHvA behavior of a 2d Fermi liquid deviates significantly from the LK formula significantly. In a 3d Fermi liquid, on the other hand, it is known that its dHvA behavior does follow the LK formula. In the literature~\cite{Luttinger1961,Gorkov-1962,maslov-2003,Nosov2024}, this has been attributed to the fact that the oscillatory part of the fermionic self energy is strongly suppressed by the small parameter $B/k_F^2$. In this section, we present a complementary analysis from the bosonized theory.

The bosonized action for a 2d Fermi liquid can be generalized to 3d~\cite{wang_unpublished}. As can be confirmed by a direct derivation, the dynamic variables $\{q,p\}$ in the zero mode sector is additionally labeled the momentum along the magnetic field, which we denote as $Z \in [-k_F,k_F]$. For a system of size $L_z$ in the $z$ direction, there are in total $N = N_{\Phi} k_F L_z / \pi$ flavors of bosons.

The oscillatory part of the grand potential is given by
\begin{align}
\Omega_{\rm osc} = -T \log&\left\{\sum_{\{p_{i,Z}\}} \exp \[-\frac{\beta\omega_c}2\sum_{i,Z}\(p_{i,Z}+\frac{\Delta_Z}{2\pi}\)^2\]\right. \nonumber\\
& ~~\times\left. \exp \[-\frac{\beta\omega_c}{2N}\sum_{i,Z}\sum_{i',Z'}\(p_{i,Z}+\frac{\Delta_Z}{2\pi}\)\(p_{i',Z'}+\frac{\Delta_Z'}{2\pi}\) F_0(Z,Z')\]\right\}.
\label{eq:3d}
\end{align}
Here each summation is over $N$ terms, and $F_0(Z,Z')$ is the zeroth Landau parameter between two different momenta along the magnetic field, i.e.,
\be
F_0(Z,Z') = \int\frac{d\theta}{2\pi} F(\theta,Z; \theta', Z') ,
\ee
where $\theta$ and $\theta'$ are azimuthal angles of two points on the Fermi surface. We note that $F_0(Z,Z')$ is not to be confused with the $F_0 = \int F(\vartheta)\, d\vartheta/(2\pi)$ usually defined for 3d Fermi liquids via Legendre polynomials, where $\vartheta$ is the angle between two Fermi wave vectors. They are generally not equal except at $Z=Z'=0$, i.e., $F_0(0,0) = F_0$.

For free fermions, $F_0(Z,Z')=0$, and the argument of the logarithm factorizes:
\begin{align}
\Omega_{\rm osc} =& -T N_{\Phi}\sum_{Z} \log\left\{\sum_{p} \exp \[-\frac{\beta\omega_c}2\(p+\frac{\Delta_Z}{2\pi}\)^2\]\right\} \non \\
=& -{T N_{\Phi}} \sum_Z \sum_{k=1}^{\infty} \frac{(-)^k \cos(k\Delta_Z)}{k\sinh(2\pi^2 k T/\omega_c)},
\end{align}
where $\Delta_Z \equiv \Delta({\pi (k_F^2-Z^2)}/{B})\in [-\pi, \pi)$ is  rapidly oscillatory as a function of $Z$.  By direct evaluation we have
\be
\sum_{Z} \cos (k\Delta_Z)  \approx L_z\int \frac{dZ}{2\pi} \cos (k\Delta_Z) = \frac{L_z}{2\pi}\sqrt{\frac{B}{k}} \cos({k \Delta_0}-\frac{\pi}{4}),
\label{eq:saddle}
\ee
where $\Delta_0= {2\pi \mathcal{A}_{\rm ext}}/{B}$ and $\mathcal{A}_{\rm ext}=\pi k_F^2$ is the area of the extremal orbit with $Z=0$. 
From this, we obtain the Lifshitz-Kosevich formula for a 3d Fermi liquid~\cite{LK1956}:
\be
\Omega_{\rm osc} = \frac{N_{\Phi}L_z T}{2\pi} \sum_{k=1}^{\infty} \frac{(-)^k \sqrt{B} \cos(k\Delta_0-\pi/4)}{k^{3/2}\sinh(2\pi^2 k T/\omega_c)}.
\label{eq:LK3d}
\ee
Heuristically, this result can be understood via a saddle point approximation around the extremal orbit, wherein the typical value of $Z$ is $\sqrt{B/k}$. As the entire range of $Z$ is $2k_F$, the rapid oscillation leads to a destructive interference when summing over $Z$ leading to a suppression ratio of $\sqrt{B}/k_F\ll 1.$

In the presence of interactions with $F_0(Z,Z')\neq 0$, the argument of the logarithm in Eq.~\eqref{eq:3d} does not factorize, making the direct evaluation of $\Omega_{\rm osc}$ more complicated. Nevertheless, from our intuition above, the typical values of  $Z,Z'$ are $\sim \sqrt{B}$, and by simple power counting, the exponent in the second line is much smaller than that in the first line by a factor of $F_0\sqrt{B}/k_F \ll 1$, in sharp contrast with the 2d case. Therefore, we expect that the leading effect of Landau parameters in 3d Fermi liquid is only via mass renormalization, and the amplitudes of dHvA follow the LK form, as shown in Eq.~\eqref{eq:LK3d}.

To demonstrated this suppression quantitatively, by analogy with Eq.~\eqref{eq:29}, we use the Poisson resummation formula on Eq.~\eqref{eq:3d} to obtain
\be
\Omega_{\rm osc} = -T \log\[\sum_{\{k_{iZ}\in \mathbb Z\}}\exp(- i \sum_{iZ} \Delta_Z k_{iZ}) \exp(-\frac{2\pi^2 T}{\omega_c}\sum_{iZ,jZ'}k_{iZ}\,\mathbb{M}^{-1}_{iZ,jZ'}\,k_{jZ'})\],
\label{eq:89}
\ee
where $\mathbb{M}$ is a $N\times N$ matrix given by
\be
\mathbb{M} = \mathbb{I}_N + \frac{\mathbb{F}_0}{N}\otimes \mathbb{J}_{N_{\Phi}},~~~\mathbb{M}^{-1} = \mathbb{I}_{N} - \mathbb{F}_0(\mathbb{I}_{N/N_{\Phi}}+\mathbb{F}_0)^{-1}\otimes\frac{\mathbb{J}_{N_{\Phi}}}{N}.
\label{eq:7.7}
\ee
Here $(\mathbb{F}_0)_{ZZ'} \equiv F_0(Z,Z')$ is a matrix of the size $\pi L_z/\pi=N/N_{\Phi}$, and  $\mathbb{I}_{N/N_{\Phi}}$ and $\mathbb{J}_{N_{\Phi}}$ are identity and all-one matrices of the specified sizes. In general, the matrix inversion in Eq.~\eqref{eq:7.7} is complicated. However, in the weak-coupling limit we can treat $\mathbb{F}_0$ perturbatively, and take
\be
\mathbb{M}^{-1} \approx \mathbb{I}_{N} - \mathbb{F}_0\otimes\frac{\mathbb{J}_{N_{\Phi}}}{N}.
\label{eq:7.7}
\ee
 
For $T\gtrsim \omega_c$ we Taylor expand $\Omega_{\rm osc}$ in terms of $\kappa=\exp(-2\pi^2T/\omega_c)\ll 1$. From a similar procedure that led to Eq.~\eqref{eq:74}, we get
\begin{align}
\Omega_{\rm osc}   -T \log&\[1+\sum_{iZ} 2\kappa \cos(\Delta_Z) \exp(\frac{\chi}{N})+ \sum_{iZ} \sum_{i' Z'\neq iZ} \kappa^2\cos(\Delta_Z + \Delta_{Z'})\exp(\frac{4\chi}{N})+\cdots\],
\label{eq:743d}
\end{align}
where as before
\be
\chi=\frac{2\pi^2T}{\omega_c}F_0.
\ee
In obtaining this we have taken the stationary phase approximation and replaced $F_0(Z,Z')$ in $\chi$ by $F_0(0,0)=F_0$, which is the usual zeroth Landau parameter in a 3d Fermi liquid~\cite{baym2008landau}.

Only keeping extensive terms, we obtain the second harmonic $\Omega_2$ by focusing on $\mathcal{O}(\kappa^2)$ terms
\begin{align}
\Omega_2=&  T\kappa^2\[ \sum_{iZ} \cos(2\Delta_Z ) -\sum_{iZ} \sum_{i' Z'} \cos(\Delta_Z + \Delta_{Z'})\frac{2\chi}{N} \].
\end{align}
Using the saddle point approximation, one can see that the first term is suppressed by $\sqrt{B}/k_F$ compared to the second due to an additional summation over $i'Z'$. By Eq.~\eqref{eq:saddle}, we get 
\be
\Omega_2 = \frac{N_{\Phi} L_z T}{2\pi} \exp(-\frac{4\pi^2 T}{\omega_c}) \sqrt{\frac{B}{2}}\[\cos(\frac{4\pi \mathcal{A}_{\rm ext}}{B}-\frac{\pi}{4}) - \sqrt{\frac{B}{k_F^2}}\frac{\pi^2T}{\omega_c}F_0\sin(\frac{4\pi \mathcal{A}_{\rm ext}}{B})\].
\ee
The first term is dominant and follows the LK formula \eqref{eq:LK3d}, while the second term leads to a parametrically small correction suppressed by $\sqrt{B}/k_F\ll 1$. This factor of suppression is consistent with the results from the fermionic theory~\cite{Luttinger1961,Gorkov-1962,maslov-2003,Nosov2024}, although to the best of our knowledge, the corrections to LK formula in 3d has not been obtained within the fermionic theory.

\section{Disorder effects and the Dingle factor }
\label{sec:dis}

Disorder effects can be treated within the bosonized theory by introducing random cosine terms $V_{ij}\cos(\phi_i - \phi_j)$ that breaks translation symmetry (see Ref.~\cite{giamarchiSchulz1988,GornyiMirlinPolyakov2007} for examples in 1d Luttinger liquids). The proper formulation and quantization of the resulting theory remain an open question. 

As we have shown, dHvA is determined completely by the zero mode sector, which is in turn controlled by single-particle properties via Eq.~\eqref{eq:single}. Instead of developing and analyzing a full-fledged bosonized field theory with disorder, we can model disorder effects in dHvA via the single-particle property of a disordered 2d Fermi liquid.

To this end, we recall that in the presence of disorder, the fermionic spectral function and the Green's function (without a magnetic field) are
\be
\rho(\omega,\vp)  = \frac{\frac{1}{2\tau}}{\[\omega-\epsilon(\vp)\]^2+\frac{1}{4\tau^2}},~~~G(\omega_m,\vp) = \int \frac{d\omega}{\pi} \frac{\rho(\omega,\vp)}{i\omega_m - \omega},
\ee
where $\tau$ is the electron-impurity scattering time.
Equivalently, the Green's function can be interpreted as one with a probabilistic chemical potential $\mu+\delta\mu$ obeying a Lorentzian distribution
\be
G(\omega_m,\vp) = \int  \frac{d \delta \mu\, p(\delta\mu)}{i\omega_m -\epsilon(\vp) + \delta\mu},~~~p(\delta \mu) = \frac{1}{\pi}\frac{\frac{1}{2\tau}}{(\delta \mu)^2+\frac{1}{4\tau^2}}.
\ee
Such a Lorentzian probability distribution of chemical potential $p(\delta \mu)$ can be directly implemented in the bosonized theory. 

With the magnetic translation symmetry broken, the $N_{\Phi}$ bosonic modes are no longer degenerate. As we discussed, we assume that for each mode the corresponding chemical potential, which enters the topological $\theta$-term, are  subject to a Lorentzian distribution. The effective  zero-mode action is 
\be
S_{\rm zero} = \int dt \sum_{i}\(-\frac{\mathcal{A}_{{\rm FS},i}}{2\pi B} p_i\dot q_i+p_i\dot q_i-\frac{ \omega_c}{2}p_i^2\) + \cdots,
\ee
where 
\be
\mathcal{A}_{{\rm FS},i} = \mathcal{A}_{\rm FS} + 2\pi m \,\delta\mu_i, 
\ee
and $m$ is the effective mass and  $\delta\mu_i$ is drawn from the Lorentzian distribution $p(\delta\mu)$. 

Through the same steps to derive the Lifshitz-Kosevich formula from the bosonized action, after  averaging $\Omega=-T\log Z$ over disorder configurations, we have
\begin{align}
\Omega_{\rm osc} =& \sum_{k}A_k\Langle\cos(\frac{k\mathcal{A}_{{\rm FS},i}}{B}) \Rangle \nonumber\\
 =& \sum_{k}A_k\int \frac{d\,\delta\mu}{\pi}\cos(\frac{k\mathcal{A}_{{\rm FS}}}{B}+\frac{2\pi k \delta \mu}{\omega_c})\frac{\frac{1}{2\tau}}{\delta\mu^2+\frac{1}{4\tau^2}} \nonumber \\
 =&  \sum_{k}A_k  \exp(-\frac{2\pi^2 k T_D }{\omega_c}) \cos(\frac{k\mathcal{A}_{{\rm FS}}}{B}),
\end{align}
where $A_k$ is the amplitude of the $k$-th dHvA harmonic in the clean case discussed in Sec.~\ref{sec:dhva}.
The extra exponential factor is known as the Dingle factor~\cite{Dingle,Brailsford}, where 
\be
T_D=\frac{1}{2\pi\tau}
\ee
is the Dingle temperature. 

Unfortunately, such a simple model cannot fully capture the disorder effects. In Refs.~\cite{maslov-2003,AdamovGornyiMirlin2006}, the authors studied the interplay between interaction and disorder, which was found to yield a temperature-dependent correction to $T_D$. To capture this feature within bosonization, one would need to introduce disorder microscopically. Moreover, impurities give rise to finite electric conductivity, which also undergoes quantum oscillations in a weak magnetic field, known as the Shubnikov-de Haas effect. However, to treat transport~\cite{Huang2024}, one would need to go beyond $S_{\rm zero}$ and couple the dynamics of all sectors by going beyond Gaussian terms in $S_{\rm LP}$. We defer these open questions to follow-up studies.

\section{Conclusion}
\label{sec:conc}
In this work, we  developed a bosonization framework for studying the dHvA effect in 2d Fermi liquids. Via the coadjoint-orbit formulation of boosnization, we circumvented the difficulties posed by the oscillatory fermionic self energy that has historically hindered an analytic solution of this problem. Our formalism naturally incorporates Landau parameters, which becomes an all-to-all coupling for $N_{\Phi}$ flavors of bosons in the presence of a weak magnetic field. We showed that the dHvA effect arises from the zero-mode sector as a topological effect, and the problem reduces to quantum mechanics in 0+1D.

We verified consistency with known Fermi liquid results by recovering Kohn’s theorem for cyclotron resonance and obtaining the linear-in-$T$ specific heat for Fermi liquids. Turning to dHvA, we derived the low- and high-temperature behaviors of the oscillation amplitudes $A_k$, showing explicit deviations from the LK formula due to interaction effects. Notably, the amplitude $A_2$ is renormalized by the Landau parameter $F_0$ in a temperature-dependent way and can even change sign at high enough temperatures. In contrast, we showed that in 3d Fermi liquids, these deviations are suppressed by $\mathcal{O}\(\sqrt{\omega_c/E_F}\)$ due to destructive interference along the direction of the field, explaining the robustness of the LK formula in three dimensions. Within the bosonic theory, we quantitatively computed the deviation from LK formula. Finally, we modeled disorder phenomenologically and recovered the expected Dingle factor.

Our results suggests possible new experimental probes based on the harmonic content of dHvA oscillations in clean 2d metals, which could serve as a diagnostic for Fermi liquid interactions. It would also be interesting to measure small deviations from the LK formula in 3d Fermi liquids and compare with our theory.
Our work opens the door to a systematic bosonic treatment of quantum oscillations in strongly correlated systems, such as non-Fermi liquids with FS's coupled to critical bosons~\cite{Guo2024,Nosov2024}. It would also be interesting to consider quantum oscillations in transport, i.e., Shubnikov-de Haas effect, which will require proper treatment of disorder and going beyond the zero-mode sector. These open questions offer fertile ground for future exploration.

\begin{acknowledgments}
I am indebted to A.~Alexandradinata, A.~Chubukov, L.~Delacretaz, D.~Else, E.~Fradkin, D.~Maslov, and P.~Nosov for enlightening discussions. I am especially grateful to M. Ye for extensive discussions and past collaboration~\cite{ye_wang_first,ye_wang_second} on this subject. This work was supported by
National Science Foundation under Grant No.~NSF DMR-2045781.  The work was partly performed at KITP at UCSB.  KITP is supported in part by the National Science Foundation under  Grant No.~NSF PHY-1748958.
\end{acknowledgments}

\bibliography{MagBosonization}

\begin{thebibliography}{65}%
\makeatletter
\providecommand \@ifxundefined [1]{%
 \@ifx{#1\undefined}
}%
\providecommand \@ifnum [1]{%
 \ifnum #1\expandafter \@firstoftwo
 \else \expandafter \@secondoftwo
 \fi
}%
\providecommand \@ifx [1]{%
 \ifx #1\expandafter \@firstoftwo
 \else \expandafter \@secondoftwo
 \fi
}%
\providecommand \natexlab [1]{#1}%
\providecommand \enquote  [1]{``#1''}%
\providecommand \bibnamefont  [1]{#1}%
\providecommand \bibfnamefont [1]{#1}%
\providecommand \citenamefont [1]{#1}%
\providecommand \href@noop [0]{\@secondoftwo}%
\providecommand \href [0]{\begingroup \@sanitize@url \@href}%
\providecommand \@href[1]{\@@startlink{#1}\@@href}%
\providecommand \@@href[1]{\endgroup#1\@@endlink}%
\providecommand \@sanitize@url [0]{\catcode `\\12\catcode `\$12\catcode
  `\&12\catcode `\#12\catcode `\^12\catcode `\_12\catcode `\%12\relax}%
\providecommand \@@startlink[1]{}%
\providecommand \@@endlink[0]{}%
\providecommand \url  [0]{\begingroup\@sanitize@url \@url }%
\providecommand \@url [1]{\endgroup\@href {#1}{\urlprefix }}%
\providecommand \urlprefix  [0]{URL }%
\providecommand \Eprint [0]{\href }%
\providecommand \doibase [0]{https://doi.org/}%
\providecommand \selectlanguage [0]{\@gobble}%
\providecommand \bibinfo  [0]{\@secondoftwo}%
\providecommand \bibfield  [0]{\@secondoftwo}%
\providecommand \translation [1]{[#1]}%
\providecommand \BibitemOpen [0]{}%
\providecommand \bibitemStop [0]{}%
\providecommand \bibitemNoStop [0]{.\EOS\space}%
\providecommand \EOS [0]{\spacefactor3000\relax}%
\providecommand \BibitemShut  [1]{\csname bibitem#1\endcsname}%
\let\auto@bib@innerbib\@empty
\bibitem [{\citenamefont {Shoenberg}(1984)}]{Shoenberg1984}%
  \BibitemOpen
  \bibfield  {author} {\bibinfo {author} {\bibfnamefont {D.}~\bibnamefont
  {Shoenberg}},\ }\href@noop {} {\emph {\bibinfo {title} {Magnetic Oscillations
  in Metals}}},\ Cambridge Monographs on Physics\ (\bibinfo  {publisher}
  {Cambridge University Press},\ \bibinfo {year} {1984})\BibitemShut {NoStop}%
\bibitem [{\citenamefont {Mikitik}\ and\ \citenamefont
  {Sharlai}(1999)}]{Mikitik1999}%
  \BibitemOpen
  \bibfield  {author} {\bibinfo {author} {\bibfnamefont {G.~P.}\ \bibnamefont
  {Mikitik}}\ and\ \bibinfo {author} {\bibfnamefont {Y.~V.}\ \bibnamefont
  {Sharlai}},\ }\bibfield  {title} {\bibinfo {title} {Manifestation of berry's
  phase in metal physics},\ }\href
  {https://doi.org/10.1103/PhysRevLett.82.2147} {\bibfield  {journal} {\bibinfo
   {journal} {Phys. Rev. Lett.}\ }\textbf {\bibinfo {volume} {82}},\ \bibinfo
  {pages} {2147} (\bibinfo {year} {1999})}\BibitemShut {NoStop}%
\bibitem [{\citenamefont {Bergemann}\ \emph {et~al.}(2000)\citenamefont
  {Bergemann}, \citenamefont {Julian}, \citenamefont {Mackenzie}, \citenamefont
  {NishiZaki},\ and\ \citenamefont {Maeno}}]{SRO-2000}%
  \BibitemOpen
  \bibfield  {author} {\bibinfo {author} {\bibfnamefont {C.}~\bibnamefont
  {Bergemann}}, \bibinfo {author} {\bibfnamefont {S.~R.}\ \bibnamefont
  {Julian}}, \bibinfo {author} {\bibfnamefont {A.~P.}\ \bibnamefont
  {Mackenzie}}, \bibinfo {author} {\bibfnamefont {S.}~\bibnamefont
  {NishiZaki}},\ and\ \bibinfo {author} {\bibfnamefont {Y.}~\bibnamefont
  {Maeno}},\ }\bibfield  {title} {\bibinfo {title} {Detailed topography of the
  fermi surface of ${\mathrm{sr}}_{2}{\mathrm{ruo}}_{4}$},\ }\href
  {https://doi.org/10.1103/PhysRevLett.84.2662} {\bibfield  {journal} {\bibinfo
   {journal} {Phys. Rev. Lett.}\ }\textbf {\bibinfo {volume} {84}},\ \bibinfo
  {pages} {2662} (\bibinfo {year} {2000})}\BibitemShut {NoStop}%
\bibitem [{\citenamefont {Mikitik}\ and\ \citenamefont
  {Sharlai}(2004)}]{mikitik-2004}%
  \BibitemOpen
  \bibfield  {author} {\bibinfo {author} {\bibfnamefont {G.~P.}\ \bibnamefont
  {Mikitik}}\ and\ \bibinfo {author} {\bibfnamefont {Y.~V.}\ \bibnamefont
  {Sharlai}},\ }\bibfield  {title} {\bibinfo {title} {Berry phase and de
  haas--van alphen effect in
  ${\mathrm{l}\mathrm{a}\mathrm{r}\mathrm{h}\mathrm{i}\mathrm{n}}_{5}$},\
  }\href {https://doi.org/10.1103/PhysRevLett.93.106403} {\bibfield  {journal}
  {\bibinfo  {journal} {Phys. Rev. Lett.}\ }\textbf {\bibinfo {volume} {93}},\
  \bibinfo {pages} {106403} (\bibinfo {year} {2004})}\BibitemShut {NoStop}%
\bibitem [{\citenamefont {Luk'yanchuk}\ and\ \citenamefont
  {Kopelevich}(2004)}]{QOgraphite2004}%
  \BibitemOpen
  \bibfield  {author} {\bibinfo {author} {\bibfnamefont {I.~A.}\ \bibnamefont
  {Luk'yanchuk}}\ and\ \bibinfo {author} {\bibfnamefont {Y.}~\bibnamefont
  {Kopelevich}},\ }\bibfield  {title} {\bibinfo {title} {Phase analysis of
  quantum oscillations in graphite},\ }\href
  {https://doi.org/10.1103/PhysRevLett.93.166402} {\bibfield  {journal}
  {\bibinfo  {journal} {Phys. Rev. Lett.}\ }\textbf {\bibinfo {volume} {93}},\
  \bibinfo {pages} {166402} (\bibinfo {year} {2004})}\BibitemShut {NoStop}%
\bibitem [{\citenamefont {Zhang}\ \emph {et~al.}(2005)\citenamefont {Zhang},
  \citenamefont {Tan}, \citenamefont {Stormer},\ and\ \citenamefont
  {Kim}}]{QOzhang2005}%
  \BibitemOpen
  \bibfield  {author} {\bibinfo {author} {\bibfnamefont {Y.}~\bibnamefont
  {Zhang}}, \bibinfo {author} {\bibfnamefont {Y.-W.}\ \bibnamefont {Tan}},
  \bibinfo {author} {\bibfnamefont {H.~L.}\ \bibnamefont {Stormer}},\ and\
  \bibinfo {author} {\bibfnamefont {P.}~\bibnamefont {Kim}},\ }\bibfield
  {title} {\bibinfo {title} {Experimental observation of the quantum hall
  effect and berry's phase in graphene},\ }\href
  {https://doi.org/10.1038/nature04235} {\bibfield  {journal} {\bibinfo
  {journal} {Nature}\ }\textbf {\bibinfo {volume} {438}},\ \bibinfo {pages}
  {201} (\bibinfo {year} {2005})}\BibitemShut {NoStop}%
\bibitem [{\citenamefont {Doiron-Leyraud}\ \emph {et~al.}(2007)\citenamefont
  {Doiron-Leyraud}, \citenamefont {Proust}, \citenamefont {LeBoeuf},
  \citenamefont {Levallois}, \citenamefont {Bonnemaison}, \citenamefont
  {Liang}, \citenamefont {Bonn}, \citenamefont {Hardy},\ and\ \citenamefont
  {Taillefer}}]{taillefer-2007}%
  \BibitemOpen
  \bibfield  {author} {\bibinfo {author} {\bibfnamefont {N.}~\bibnamefont
  {Doiron-Leyraud}}, \bibinfo {author} {\bibfnamefont {C.}~\bibnamefont
  {Proust}}, \bibinfo {author} {\bibfnamefont {D.}~\bibnamefont {LeBoeuf}},
  \bibinfo {author} {\bibfnamefont {J.}~\bibnamefont {Levallois}}, \bibinfo
  {author} {\bibfnamefont {J.-B.}\ \bibnamefont {Bonnemaison}}, \bibinfo
  {author} {\bibfnamefont {R.}~\bibnamefont {Liang}}, \bibinfo {author}
  {\bibfnamefont {D.~A.}\ \bibnamefont {Bonn}}, \bibinfo {author}
  {\bibfnamefont {W.~N.}\ \bibnamefont {Hardy}},\ and\ \bibinfo {author}
  {\bibfnamefont {L.}~\bibnamefont {Taillefer}},\ }\bibfield  {title} {\bibinfo
  {title} {Quantum oscillations and the fermi surface in an underdoped high-tc
  superconductor},\ }\href {https://doi.org/10.1038/nature05872} {\bibfield
  {journal} {\bibinfo  {journal} {Nature}\ }\textbf {\bibinfo {volume} {447}},\
  \bibinfo {pages} {565–568} (\bibinfo {year} {2007})}\BibitemShut {NoStop}%
\bibitem [{\citenamefont {VanGennep}\ \emph {et~al.}(2014)\citenamefont
  {VanGennep}, \citenamefont {Maiti}, \citenamefont {Graf}, \citenamefont
  {Tozer}, \citenamefont {Martin}, \citenamefont {Berger}, \citenamefont
  {Maslov},\ and\ \citenamefont {Hamlin}}]{maslov-hamlin-2014}%
  \BibitemOpen
  \bibfield  {author} {\bibinfo {author} {\bibfnamefont {D.}~\bibnamefont
  {VanGennep}}, \bibinfo {author} {\bibfnamefont {S.}~\bibnamefont {Maiti}},
  \bibinfo {author} {\bibfnamefont {D.}~\bibnamefont {Graf}}, \bibinfo {author}
  {\bibfnamefont {S.~W.}\ \bibnamefont {Tozer}}, \bibinfo {author}
  {\bibfnamefont {C.}~\bibnamefont {Martin}}, \bibinfo {author} {\bibfnamefont
  {H.}~\bibnamefont {Berger}}, \bibinfo {author} {\bibfnamefont {D.~L.}\
  \bibnamefont {Maslov}},\ and\ \bibinfo {author} {\bibfnamefont {J.~J.}\
  \bibnamefont {Hamlin}},\ }\bibfield  {title} {\bibinfo {title} {Pressure
  tuning the fermi level through the dirac point of giant rashba semiconductor
  bitei},\ }\href {https://doi.org/10.1088/0953-8984/26/34/342202} {\bibfield
  {journal} {\bibinfo  {journal} {Journal of Physics: Condensed Matter}\
  }\textbf {\bibinfo {volume} {26}},\ \bibinfo {pages} {342202} (\bibinfo
  {year} {2014})}\BibitemShut {NoStop}%
\bibitem [{\citenamefont {Sebastian}\ and\ \citenamefont
  {Proust}(2015)}]{Sebastian-2015}%
  \BibitemOpen
  \bibfield  {author} {\bibinfo {author} {\bibfnamefont {S.~E.}\ \bibnamefont
  {Sebastian}}\ and\ \bibinfo {author} {\bibfnamefont {C.}~\bibnamefont
  {Proust}},\ }\bibfield  {title} {\bibinfo {title} {Quantum oscillations in
  hole-doped cuprates},\ }\href
  {https://doi.org/10.1146/annurev-conmatphys-030212-184305} {\bibfield
  {journal} {\bibinfo  {journal} {Annual Review of Condensed Matter Physics}\
  }\textbf {\bibinfo {volume} {6}},\ \bibinfo {pages} {411–430} (\bibinfo
  {year} {2015})}\BibitemShut {NoStop}%
\bibitem [{\citenamefont {Tan}\ \emph {et~al.}(2015)\citenamefont {Tan},
  \citenamefont {Hsu}, \citenamefont {Zeng}, \citenamefont {Hatnean},
  \citenamefont {Harrison}, \citenamefont {Zhu}, \citenamefont {Hartstein},
  \citenamefont {Kiourlappou}, \citenamefont {Srivastava}, \citenamefont
  {Johannes}, \citenamefont {Murphy}, \citenamefont {Park}, \citenamefont
  {Balicas}, \citenamefont {Lonzarich}, \citenamefont {Balakrishnan},\ and\
  \citenamefont {Sebastian}}]{Sebastian-SmB6-2015}%
  \BibitemOpen
  \bibfield  {author} {\bibinfo {author} {\bibfnamefont {B.~S.}\ \bibnamefont
  {Tan}}, \bibinfo {author} {\bibfnamefont {Y.-T.}\ \bibnamefont {Hsu}},
  \bibinfo {author} {\bibfnamefont {B.}~\bibnamefont {Zeng}}, \bibinfo {author}
  {\bibfnamefont {M.~C.}\ \bibnamefont {Hatnean}}, \bibinfo {author}
  {\bibfnamefont {N.}~\bibnamefont {Harrison}}, \bibinfo {author}
  {\bibfnamefont {Z.}~\bibnamefont {Zhu}}, \bibinfo {author} {\bibfnamefont
  {M.}~\bibnamefont {Hartstein}}, \bibinfo {author} {\bibfnamefont
  {M.}~\bibnamefont {Kiourlappou}}, \bibinfo {author} {\bibfnamefont
  {A.}~\bibnamefont {Srivastava}}, \bibinfo {author} {\bibfnamefont {M.~D.}\
  \bibnamefont {Johannes}}, \bibinfo {author} {\bibfnamefont {T.~P.}\
  \bibnamefont {Murphy}}, \bibinfo {author} {\bibfnamefont {J.-H.}\
  \bibnamefont {Park}}, \bibinfo {author} {\bibfnamefont {L.}~\bibnamefont
  {Balicas}}, \bibinfo {author} {\bibfnamefont {G.~G.}\ \bibnamefont
  {Lonzarich}}, \bibinfo {author} {\bibfnamefont {G.}~\bibnamefont
  {Balakrishnan}},\ and\ \bibinfo {author} {\bibfnamefont {S.~E.}\ \bibnamefont
  {Sebastian}},\ }\bibfield  {title} {\bibinfo {title} {Unconventional fermi
  surface in an insulating state},\ }\href
  {https://doi.org/10.1126/science.aaa7974} {\bibfield  {journal} {\bibinfo
  {journal} {Science}\ }\textbf {\bibinfo {volume} {349}},\ \bibinfo {pages}
  {287–290} (\bibinfo {year} {2015})}\BibitemShut {NoStop}%
\bibitem [{\citenamefont {Murakawa}\ \emph {et~al.}(2013)\citenamefont
  {Murakawa}, \citenamefont {Bahramy}, \citenamefont {Tokunaga}, \citenamefont
  {Kohama}, \citenamefont {Bell}, \citenamefont {Kaneko}, \citenamefont
  {Nagaosa}, \citenamefont {Hwang},\ and\ \citenamefont
  {Tokura}}]{Berry-Rashba}%
  \BibitemOpen
  \bibfield  {author} {\bibinfo {author} {\bibfnamefont {H.}~\bibnamefont
  {Murakawa}}, \bibinfo {author} {\bibfnamefont {M.~S.}\ \bibnamefont
  {Bahramy}}, \bibinfo {author} {\bibfnamefont {M.}~\bibnamefont {Tokunaga}},
  \bibinfo {author} {\bibfnamefont {Y.}~\bibnamefont {Kohama}}, \bibinfo
  {author} {\bibfnamefont {C.}~\bibnamefont {Bell}}, \bibinfo {author}
  {\bibfnamefont {Y.}~\bibnamefont {Kaneko}}, \bibinfo {author} {\bibfnamefont
  {N.}~\bibnamefont {Nagaosa}}, \bibinfo {author} {\bibfnamefont {H.~Y.}\
  \bibnamefont {Hwang}},\ and\ \bibinfo {author} {\bibfnamefont
  {Y.}~\bibnamefont {Tokura}},\ }\bibfield  {title} {\bibinfo {title}
  {Detection of berry’s phase in a bulk rashba semiconductor},\ }\href
  {https://doi.org/10.1126/science.1242247} {\bibfield  {journal} {\bibinfo
  {journal} {Science}\ }\textbf {\bibinfo {volume} {342}},\ \bibinfo {pages}
  {1490} (\bibinfo {year} {2013})},\ \Eprint
  {https://arxiv.org/abs/https://www.science.org/doi/pdf/10.1126/science.1242247}
  {https://www.science.org/doi/pdf/10.1126/science.1242247} \BibitemShut
  {NoStop}%
\bibitem [{\citenamefont {Pariari}\ \emph {et~al.}(2015)\citenamefont
  {Pariari}, \citenamefont {Dutta},\ and\ \citenamefont
  {Mandal}}]{cd3as2-dhva}%
  \BibitemOpen
  \bibfield  {author} {\bibinfo {author} {\bibfnamefont {A.}~\bibnamefont
  {Pariari}}, \bibinfo {author} {\bibfnamefont {P.}~\bibnamefont {Dutta}},\
  and\ \bibinfo {author} {\bibfnamefont {P.}~\bibnamefont {Mandal}},\
  }\bibfield  {title} {\bibinfo {title} {Probing the fermi surface of
  three-dimensional dirac semimetal ${\mathrm{cd}}_{3}{\mathrm{as}}_{2}$
  through the de haas--van alphen technique},\ }\href
  {https://doi.org/10.1103/PhysRevB.91.155139} {\bibfield  {journal} {\bibinfo
  {journal} {Phys. Rev. B}\ }\textbf {\bibinfo {volume} {91}},\ \bibinfo
  {pages} {155139} (\bibinfo {year} {2015})}\BibitemShut {NoStop}%
\bibitem [{\citenamefont {Wu}\ \emph {et~al.}(2019)\citenamefont {Wu},
  \citenamefont {Guo}, \citenamefont {Smidman}, \citenamefont {Zhang},
  \citenamefont {Chen}, \citenamefont {Singleton},\ and\ \citenamefont
  {Yuan}}]{wu2019anomalous}%
  \BibitemOpen
  \bibfield  {author} {\bibinfo {author} {\bibfnamefont {F.}~\bibnamefont
  {Wu}}, \bibinfo {author} {\bibfnamefont {C.}~\bibnamefont {Guo}}, \bibinfo
  {author} {\bibfnamefont {M.}~\bibnamefont {Smidman}}, \bibinfo {author}
  {\bibfnamefont {J.}~\bibnamefont {Zhang}}, \bibinfo {author} {\bibfnamefont
  {Y.}~\bibnamefont {Chen}}, \bibinfo {author} {\bibfnamefont {J.}~\bibnamefont
  {Singleton}},\ and\ \bibinfo {author} {\bibfnamefont {H.}~\bibnamefont
  {Yuan}},\ }\bibfield  {title} {\bibinfo {title} {Anomalous quantum
  oscillations and evidence for a non-trivial berry phase in smsb},\
  }\href@noop {} {\bibfield  {journal} {\bibinfo  {journal} {npj Quantum
  Materials}\ }\textbf {\bibinfo {volume} {4}},\ \bibinfo {pages} {20}
  (\bibinfo {year} {2019})}\BibitemShut {NoStop}%
\bibitem [{\citenamefont {Ye}\ \emph {et~al.}(2019)\citenamefont {Ye},
  \citenamefont {Chan}, \citenamefont {McDonald}, \citenamefont {Graf},
  \citenamefont {Kang}, \citenamefont {Liu}, \citenamefont {Suzuki},
  \citenamefont {Comin}, \citenamefont {Fu},\ and\ \citenamefont
  {Checkelsky}}]{ye2019haas}%
  \BibitemOpen
  \bibfield  {author} {\bibinfo {author} {\bibfnamefont {L.}~\bibnamefont
  {Ye}}, \bibinfo {author} {\bibfnamefont {M.~K.}\ \bibnamefont {Chan}},
  \bibinfo {author} {\bibfnamefont {R.~D.}\ \bibnamefont {McDonald}}, \bibinfo
  {author} {\bibfnamefont {D.}~\bibnamefont {Graf}}, \bibinfo {author}
  {\bibfnamefont {M.}~\bibnamefont {Kang}}, \bibinfo {author} {\bibfnamefont
  {J.}~\bibnamefont {Liu}}, \bibinfo {author} {\bibfnamefont {T.}~\bibnamefont
  {Suzuki}}, \bibinfo {author} {\bibfnamefont {R.}~\bibnamefont {Comin}},
  \bibinfo {author} {\bibfnamefont {L.}~\bibnamefont {Fu}},\ and\ \bibinfo
  {author} {\bibfnamefont {J.~G.}\ \bibnamefont {Checkelsky}},\ }\bibfield
  {title} {\bibinfo {title} {de haas-van alphen effect of correlated dirac
  states in kagome metal fe3sn2},\ }\href@noop {} {\bibfield  {journal}
  {\bibinfo  {journal} {Nature communications}\ }\textbf {\bibinfo {volume}
  {10}},\ \bibinfo {pages} {4870} (\bibinfo {year} {2019})}\BibitemShut
  {NoStop}%
\bibitem [{\citenamefont {Alexandradinata}\ and\ \citenamefont
  {Glazman}(2023)}]{Alexandradinata2023ar}%
  \BibitemOpen
  \bibfield  {author} {\bibinfo {author} {\bibfnamefont {A.}~\bibnamefont
  {Alexandradinata}}\ and\ \bibinfo {author} {\bibfnamefont {L.}~\bibnamefont
  {Glazman}},\ }\bibfield  {title} {\bibinfo {title} {Fermiology of topological
  metals},\ }\href
  {https://doi.org/https://doi.org/10.1146/annurev-conmatphys-040721-021331}
  {\bibfield  {journal} {\bibinfo  {journal} {Annual Review of Condensed Matter
  Physics}\ }\textbf {\bibinfo {volume} {14}},\ \bibinfo {pages} {261}
  (\bibinfo {year} {2023})}\BibitemShut {NoStop}%
\bibitem [{\citenamefont {Shrestha}\ \emph {et~al.}(2022)\citenamefont
  {Shrestha}, \citenamefont {Chapai}, \citenamefont {Pokharel}, \citenamefont
  {Miertschin}, \citenamefont {Nguyen}, \citenamefont {Zhou}, \citenamefont
  {Chung}, \citenamefont {Kanatzidis}, \citenamefont {Mitchell}, \citenamefont
  {Welp}, \citenamefont {Popovi\ifmmode~\acute{c}\else \'{c}\fi{}},
  \citenamefont {Graf}, \citenamefont {Lorenz},\ and\ \citenamefont
  {Kwok}}]{QOkagome2022}%
  \BibitemOpen
  \bibfield  {author} {\bibinfo {author} {\bibfnamefont {K.}~\bibnamefont
  {Shrestha}}, \bibinfo {author} {\bibfnamefont {R.}~\bibnamefont {Chapai}},
  \bibinfo {author} {\bibfnamefont {B.~K.}\ \bibnamefont {Pokharel}}, \bibinfo
  {author} {\bibfnamefont {D.}~\bibnamefont {Miertschin}}, \bibinfo {author}
  {\bibfnamefont {T.}~\bibnamefont {Nguyen}}, \bibinfo {author} {\bibfnamefont
  {X.}~\bibnamefont {Zhou}}, \bibinfo {author} {\bibfnamefont {D.~Y.}\
  \bibnamefont {Chung}}, \bibinfo {author} {\bibfnamefont {M.~G.}\ \bibnamefont
  {Kanatzidis}}, \bibinfo {author} {\bibfnamefont {J.~F.}\ \bibnamefont
  {Mitchell}}, \bibinfo {author} {\bibfnamefont {U.}~\bibnamefont {Welp}},
  \bibinfo {author} {\bibfnamefont {D.}~\bibnamefont
  {Popovi\ifmmode~\acute{c}\else \'{c}\fi{}}}, \bibinfo {author} {\bibfnamefont
  {D.~E.}\ \bibnamefont {Graf}}, \bibinfo {author} {\bibfnamefont
  {B.}~\bibnamefont {Lorenz}},\ and\ \bibinfo {author} {\bibfnamefont {W.~K.}\
  \bibnamefont {Kwok}},\ }\bibfield  {title} {\bibinfo {title} {Nontrivial
  fermi surface topology of the kagome superconductor
  ${\mathrm{csv}}_{3}{\mathrm{sb}}_{5}$ probed by de haas--van alphen
  oscillations},\ }\href {https://doi.org/10.1103/PhysRevB.105.024508}
  {\bibfield  {journal} {\bibinfo  {journal} {Phys. Rev. B}\ }\textbf {\bibinfo
  {volume} {105}},\ \bibinfo {pages} {024508} (\bibinfo {year}
  {2022})}\BibitemShut {NoStop}%
\bibitem [{\citenamefont {Chapai}\ \emph {et~al.}(2023)\citenamefont {Chapai},
  \citenamefont {Leroux}, \citenamefont {Oliviero}, \citenamefont {Vignolles},
  \citenamefont {Bruyant}, \citenamefont {Smylie}, \citenamefont {Chung},
  \citenamefont {Kanatzidis}, \citenamefont {Kwok}, \citenamefont {Mitchell},\
  and\ \citenamefont {Welp}}]{QOkagome2023}%
  \BibitemOpen
  \bibfield  {author} {\bibinfo {author} {\bibfnamefont {R.}~\bibnamefont
  {Chapai}}, \bibinfo {author} {\bibfnamefont {M.}~\bibnamefont {Leroux}},
  \bibinfo {author} {\bibfnamefont {V.}~\bibnamefont {Oliviero}}, \bibinfo
  {author} {\bibfnamefont {D.}~\bibnamefont {Vignolles}}, \bibinfo {author}
  {\bibfnamefont {N.}~\bibnamefont {Bruyant}}, \bibinfo {author} {\bibfnamefont
  {M.~P.}\ \bibnamefont {Smylie}}, \bibinfo {author} {\bibfnamefont {D.~Y.}\
  \bibnamefont {Chung}}, \bibinfo {author} {\bibfnamefont {M.~G.}\ \bibnamefont
  {Kanatzidis}}, \bibinfo {author} {\bibfnamefont {W.-K.}\ \bibnamefont
  {Kwok}}, \bibinfo {author} {\bibfnamefont {J.~F.}\ \bibnamefont {Mitchell}},\
  and\ \bibinfo {author} {\bibfnamefont {U.}~\bibnamefont {Welp}},\ }\bibfield
  {title} {\bibinfo {title} {Magnetic breakdown and topology in the kagome
  superconductor ${\mathrm{csv}}_{3}{\mathrm{sb}}_{5}$ under high magnetic
  field},\ }\href {https://doi.org/10.1103/PhysRevLett.130.126401} {\bibfield
  {journal} {\bibinfo  {journal} {Phys. Rev. Lett.}\ }\textbf {\bibinfo
  {volume} {130}},\ \bibinfo {pages} {126401} (\bibinfo {year}
  {2023})}\BibitemShut {NoStop}%
\bibitem [{\citenamefont {Li}\ \emph {et~al.}(2023)\citenamefont {Li},
  \citenamefont {Tan},\ and\ \citenamefont
  {Yan}}]{li2023quantumoscillationstopologicalphases}%
  \BibitemOpen
  \bibfield  {author} {\bibinfo {author} {\bibfnamefont {Y.}~\bibnamefont
  {Li}}, \bibinfo {author} {\bibfnamefont {H.}~\bibnamefont {Tan}},\ and\
  \bibinfo {author} {\bibfnamefont {B.}~\bibnamefont {Yan}},\ }\href
  {https://arxiv.org/abs/2307.04750} {\bibinfo {title} {Quantum oscillations
  with topological phases in a kagome metal csti$_3$bi$_5$}} (\bibinfo {year}
  {2023}),\ \Eprint {https://arxiv.org/abs/2307.04750} {arXiv:2307.04750
  [cond-mat.str-el]} \BibitemShut {NoStop}%
\bibitem [{\citenamefont {Guo}\ \emph {et~al.}(2024)\citenamefont {Guo},
  \citenamefont {Valentinis}, \citenamefont {Schmalian}, \citenamefont
  {Sachdev},\ and\ \citenamefont {Patel}}]{Guo2024}%
  \BibitemOpen
  \bibfield  {author} {\bibinfo {author} {\bibfnamefont {H.}~\bibnamefont
  {Guo}}, \bibinfo {author} {\bibfnamefont {D.}~\bibnamefont {Valentinis}},
  \bibinfo {author} {\bibfnamefont {J.}~\bibnamefont {Schmalian}}, \bibinfo
  {author} {\bibfnamefont {S.}~\bibnamefont {Sachdev}},\ and\ \bibinfo {author}
  {\bibfnamefont {A.~A.}\ \bibnamefont {Patel}},\ }\bibfield  {title} {\bibinfo
  {title} {Cyclotron resonance and quantum oscillations of critical fermi
  surfaces},\ }\href {https://doi.org/10.1103/PhysRevB.109.075162} {\bibfield
  {journal} {\bibinfo  {journal} {Phys. Rev. B}\ }\textbf {\bibinfo {volume}
  {109}},\ \bibinfo {pages} {075162} (\bibinfo {year} {2024})}\BibitemShut
  {NoStop}%
\bibitem [{\citenamefont {Lifshitz}\ and\ \citenamefont
  {Kosevich}(1956)}]{LK1956}%
  \BibitemOpen
  \bibfield  {author} {\bibinfo {author} {\bibfnamefont {I.}~\bibnamefont
  {Lifshitz}}\ and\ \bibinfo {author} {\bibfnamefont {A.~M.}\ \bibnamefont
  {Kosevich}},\ }\bibfield  {title} {\bibinfo {title} {Theory of magnetic
  susceptibility in metals at low temperatures},\ }\href@noop {} {\bibfield
  {journal} {\bibinfo  {journal} {Sov. Phys. JETP}\ }\textbf {\bibinfo {volume}
  {2}},\ \bibinfo {pages} {636} (\bibinfo {year} {1956})}\BibitemShut {NoStop}%
\bibitem [{\citenamefont {Chang}\ and\ \citenamefont
  {Niu}(1996)}]{chang-niu-1996}%
  \BibitemOpen
  \bibfield  {author} {\bibinfo {author} {\bibfnamefont {M.-C.}\ \bibnamefont
  {Chang}}\ and\ \bibinfo {author} {\bibfnamefont {Q.}~\bibnamefont {Niu}},\
  }\bibfield  {title} {\bibinfo {title} {Berry phase, hyperorbits, and the
  hofstadter spectrum: Semiclassical dynamics in magnetic bloch bands},\ }\href
  {https://doi.org/10.1103/PhysRevB.53.7010} {\bibfield  {journal} {\bibinfo
  {journal} {Phys. Rev. B}\ }\textbf {\bibinfo {volume} {53}},\ \bibinfo
  {pages} {7010} (\bibinfo {year} {1996})}\BibitemShut {NoStop}%
\bibitem [{\citenamefont {Ye}\ and\ \citenamefont
  {Wang}(2025)}]{ye_wang_second}%
  \BibitemOpen
  \bibfield  {author} {\bibinfo {author} {\bibfnamefont {M.}~\bibnamefont
  {Ye}}\ and\ \bibinfo {author} {\bibfnamefont {Y.}~\bibnamefont {Wang}},\
  }\href {https://arxiv.org/abs/2412.16289} {\bibinfo {title} {Berry phase and
  quantum oscillation from multi-orbital coadjoint-orbit bosonization}}
  (\bibinfo {year} {2025}),\ \Eprint {https://arxiv.org/abs/2412.16289}
  {arXiv:2412.16289 [cond-mat.str-el]} \BibitemShut {NoStop}%
\bibitem [{\citenamefont {Luttinger}(1961)}]{Luttinger1961}%
  \BibitemOpen
  \bibfield  {author} {\bibinfo {author} {\bibfnamefont {J.~M.}\ \bibnamefont
  {Luttinger}},\ }\bibfield  {title} {\bibinfo {title} {Theory of the de
  haas-van alphen effect for a system of interacting fermions},\ }\href
  {https://doi.org/10.1103/PhysRev.121.1251} {\bibfield  {journal} {\bibinfo
  {journal} {Phys. Rev.}\ }\textbf {\bibinfo {volume} {121}},\ \bibinfo {pages}
  {1251} (\bibinfo {year} {1961})}\BibitemShut {NoStop}%
\bibitem [{\citenamefont {Gor’kov}\ and\ \citenamefont
  {Bychkov}(1962)}]{Gorkov-1962}%
  \BibitemOpen
  \bibfield  {author} {\bibinfo {author} {\bibfnamefont {L.}~\bibnamefont
  {Gor’kov}}\ and\ \bibinfo {author} {\bibfnamefont {Y.~A.}\ \bibnamefont
  {Bychkov}},\ }\bibfield  {title} {\bibinfo {title} {Quantum oscillations of
  the thermodynamic quantities of a metal in a magnetic field according to the
  fermi-liquid model},\ }\href@noop {} {\bibfield  {journal} {\bibinfo
  {journal} {JETP}\ }\textbf {\bibinfo {volume} {14}} (\bibinfo {year}
  {1962})}\BibitemShut {NoStop}%
\bibitem [{\citenamefont {Martin}\ \emph {et~al.}(2003)\citenamefont {Martin},
  \citenamefont {Maslov},\ and\ \citenamefont {Reizer}}]{maslov-2003}%
  \BibitemOpen
  \bibfield  {author} {\bibinfo {author} {\bibfnamefont {G.~W.}\ \bibnamefont
  {Martin}}, \bibinfo {author} {\bibfnamefont {D.~L.}\ \bibnamefont {Maslov}},\
  and\ \bibinfo {author} {\bibfnamefont {M.~Y.}\ \bibnamefont {Reizer}},\
  }\bibfield  {title} {\bibinfo {title} {Quantum magneto-oscillations in a
  two-dimensional fermi liquid},\ }\href
  {https://doi.org/10.1103/PhysRevB.68.241309} {\bibfield  {journal} {\bibinfo
  {journal} {Phys. Rev. B}\ }\textbf {\bibinfo {volume} {68}},\ \bibinfo
  {pages} {241309} (\bibinfo {year} {2003})}\BibitemShut {NoStop}%
\bibitem [{\citenamefont {Nosov}\ \emph {et~al.}(2024)\citenamefont {Nosov},
  \citenamefont {Wu},\ and\ \citenamefont {Raghu}}]{Nosov2024}%
  \BibitemOpen
  \bibfield  {author} {\bibinfo {author} {\bibfnamefont {P.~A.}\ \bibnamefont
  {Nosov}}, \bibinfo {author} {\bibfnamefont {Y.-M.}\ \bibnamefont {Wu}},\ and\
  \bibinfo {author} {\bibfnamefont {S.}~\bibnamefont {Raghu}},\ }\bibfield
  {title} {\bibinfo {title} {Entropy and de haas--van alphen oscillations of a
  three-dimensional marginal fermi liquid},\ }\href
  {https://doi.org/10.1103/PhysRevB.109.075107} {\bibfield  {journal} {\bibinfo
   {journal} {Phys. Rev. B}\ }\textbf {\bibinfo {volume} {109}},\ \bibinfo
  {pages} {075107} (\bibinfo {year} {2024})}\BibitemShut {NoStop}%
\bibitem [{\citenamefont {Koulakov}\ \emph {et~al.}(1996)\citenamefont
  {Koulakov}, \citenamefont {Fogler},\ and\ \citenamefont
  {Shklovskii}}]{koulakov1996chargedensity}%
  \BibitemOpen
  \bibfield  {author} {\bibinfo {author} {\bibfnamefont {A.~A.}\ \bibnamefont
  {Koulakov}}, \bibinfo {author} {\bibfnamefont {M.~M.}\ \bibnamefont
  {Fogler}},\ and\ \bibinfo {author} {\bibfnamefont {B.~I.}\ \bibnamefont
  {Shklovskii}},\ }\bibfield  {title} {\bibinfo {title} {Charge density wave in
  two‐dimensional electron liquid in weak magnetic field},\ }\href
  {https://doi.org/10.1103/PhysRevLett.76.499} {\bibfield  {journal} {\bibinfo
  {journal} {Physical Review Letters}\ }\textbf {\bibinfo {volume} {76}},\
  \bibinfo {pages} {499} (\bibinfo {year} {1996})}\BibitemShut {NoStop}%
\bibitem [{\citenamefont {Lilly}\ \emph {et~al.}(1999)\citenamefont {Lilly},
  \citenamefont {Cooper}, \citenamefont {Eisenstein}, \citenamefont
  {Pfeiffer},\ and\ \citenamefont {West}}]{lilly1999anisotropic}%
  \BibitemOpen
  \bibfield  {author} {\bibinfo {author} {\bibfnamefont {M.~P.}\ \bibnamefont
  {Lilly}}, \bibinfo {author} {\bibfnamefont {K.~B.}\ \bibnamefont {Cooper}},
  \bibinfo {author} {\bibfnamefont {J.~P.}\ \bibnamefont {Eisenstein}},
  \bibinfo {author} {\bibfnamefont {L.~N.}\ \bibnamefont {Pfeiffer}},\ and\
  \bibinfo {author} {\bibfnamefont {K.~W.}\ \bibnamefont {West}},\ }\bibfield
  {title} {\bibinfo {title} {Evidence for an anisotropic state of
  two‐dimensional electrons in high landau levels},\ }\href
  {https://doi.org/10.1103/PhysRevLett.82.394} {\bibfield  {journal} {\bibinfo
  {journal} {Physical Review Letters}\ }\textbf {\bibinfo {volume} {82}},\
  \bibinfo {pages} {394} (\bibinfo {year} {1999})}\BibitemShut {NoStop}%
\bibitem [{\citenamefont {Pan}\ \emph {et~al.}(1999)\citenamefont {Pan},
  \citenamefont {Du}, \citenamefont {Stormer}, \citenamefont {Tsui},
  \citenamefont {Pfeiffer}, \citenamefont {Baldwin},\ and\ \citenamefont
  {West}}]{pan1999anisotropic}%
  \BibitemOpen
  \bibfield  {author} {\bibinfo {author} {\bibfnamefont {W.}~\bibnamefont
  {Pan}}, \bibinfo {author} {\bibfnamefont {R.~R.}\ \bibnamefont {Du}},
  \bibinfo {author} {\bibfnamefont {H.~L.}\ \bibnamefont {Stormer}}, \bibinfo
  {author} {\bibfnamefont {D.~C.}\ \bibnamefont {Tsui}}, \bibinfo {author}
  {\bibfnamefont {L.~N.}\ \bibnamefont {Pfeiffer}}, \bibinfo {author}
  {\bibfnamefont {K.~W.}\ \bibnamefont {Baldwin}},\ and\ \bibinfo {author}
  {\bibfnamefont {K.~W.}\ \bibnamefont {West}},\ }\bibfield  {title} {\bibinfo
  {title} {Strongly anisotropic electronic transport at landau level filling
  factor $\nu = 9/2$ and $\nu = 5/2$ under a tilted magnetic field},\ }\href
  {https://doi.org/10.1103/PhysRevLett.83.820} {\bibfield  {journal} {\bibinfo
  {journal} {Physical Review Letters}\ }\textbf {\bibinfo {volume} {83}},\
  \bibinfo {pages} {820} (\bibinfo {year} {1999})}\BibitemShut {NoStop}%
\bibitem [{\citenamefont {Stormer}\ \emph {et~al.}(1999)\citenamefont
  {Stormer}, \citenamefont {Tsui},\ and\ \citenamefont
  {Gossard}}]{Stormer1999}%
  \BibitemOpen
  \bibfield  {author} {\bibinfo {author} {\bibfnamefont {H.~L.}\ \bibnamefont
  {Stormer}}, \bibinfo {author} {\bibfnamefont {D.~C.}\ \bibnamefont {Tsui}},\
  and\ \bibinfo {author} {\bibfnamefont {A.~C.}\ \bibnamefont {Gossard}},\
  }\bibfield  {title} {\bibinfo {title} {The fractional quantum hall effect},\
  }\href {https://doi.org/10.1103/RevModPhys.71.S298} {\bibfield  {journal}
  {\bibinfo  {journal} {Rev. Mod. Phys.}\ }\textbf {\bibinfo {volume} {71}},\
  \bibinfo {pages} {S298} (\bibinfo {year} {1999})}\BibitemShut {NoStop}%
\bibitem [{\citenamefont {Miyake}\ and\ \citenamefont
  {Varma}(1993)}]{Miyake1993}%
  \BibitemOpen
  \bibfield  {author} {\bibinfo {author} {\bibfnamefont {K.}~\bibnamefont
  {Miyake}}\ and\ \bibinfo {author} {\bibfnamefont {C.}~\bibnamefont {Varma}},\
  }\bibfield  {title} {\bibinfo {title} {Many body effect on oscillatory
  properties of two-dimensional metals in a magnetic field},\ }\href
  {https://doi.org/https://doi.org/10.1016/0038-1098(93)90027-K} {\bibfield
  {journal} {\bibinfo  {journal} {Solid State Communications}\ }\textbf
  {\bibinfo {volume} {85}},\ \bibinfo {pages} {335} (\bibinfo {year}
  {1993})}\BibitemShut {NoStop}%
\bibitem [{\citenamefont {Curnoe}\ and\ \citenamefont
  {Stamp}(1998)}]{CurnoeStamp1998}%
  \BibitemOpen
  \bibfield  {author} {\bibinfo {author} {\bibfnamefont {S.}~\bibnamefont
  {Curnoe}}\ and\ \bibinfo {author} {\bibfnamefont {P.~C.~E.}\ \bibnamefont
  {Stamp}},\ }\bibfield  {title} {\bibinfo {title} {Quantum oscillations of
  electrons and of composite fermions in two dimensions: Beyond the luttinger
  expansion},\ }\href {https://doi.org/10.1103/PhysRevLett.80.3312} {\bibfield
  {journal} {\bibinfo  {journal} {Phys. Rev. Lett.}\ }\textbf {\bibinfo
  {volume} {80}},\ \bibinfo {pages} {3312} (\bibinfo {year}
  {1998})}\BibitemShut {NoStop}%
\bibitem [{\citenamefont {Haldane}(2005)}]{HaldaneBosonization2005}%
  \BibitemOpen
  \bibfield  {author} {\bibinfo {author} {\bibfnamefont {F.~D.~M.}\
  \bibnamefont {Haldane}},\ }\bibfield  {title} {\bibinfo {title} {Luttinger's
  theorem and bosonization of the fermi surface}\ }\href
  {https://doi.org/10.48550/ARXIV.COND-MAT/0505529}
  {10.48550/ARXIV.COND-MAT/0505529} (\bibinfo {year} {2005})\BibitemShut
  {NoStop}%
\bibitem [{\citenamefont {Castro~Neto}\ and\ \citenamefont
  {Fradkin}(1994{\natexlab{a}})}]{FradkinBosonization1994}%
  \BibitemOpen
  \bibfield  {author} {\bibinfo {author} {\bibfnamefont {A.~H.}\ \bibnamefont
  {Castro~Neto}}\ and\ \bibinfo {author} {\bibfnamefont {E.}~\bibnamefont
  {Fradkin}},\ }\bibfield  {title} {\bibinfo {title} {Bosonization of fermi
  liquids},\ }\href {https://doi.org/10.1103/PhysRevB.49.10877} {\bibfield
  {journal} {\bibinfo  {journal} {Phys. Rev. B}\ }\textbf {\bibinfo {volume}
  {49}},\ \bibinfo {pages} {10877} (\bibinfo {year}
  {1994}{\natexlab{a}})}\BibitemShut {NoStop}%
\bibitem [{\citenamefont {Castro~Neto}\ and\ \citenamefont
  {Fradkin}(1994{\natexlab{b}})}]{FradkinBosonizationPRL}%
  \BibitemOpen
  \bibfield  {author} {\bibinfo {author} {\bibfnamefont {A.~H.}\ \bibnamefont
  {Castro~Neto}}\ and\ \bibinfo {author} {\bibfnamefont {E.}~\bibnamefont
  {Fradkin}},\ }\bibfield  {title} {\bibinfo {title} {Bosonization of the low
  energy excitations of fermi liquids},\ }\href
  {https://doi.org/10.1103/PhysRevLett.72.1393} {\bibfield  {journal} {\bibinfo
   {journal} {Phys. Rev. Lett.}\ }\textbf {\bibinfo {volume} {72}},\ \bibinfo
  {pages} {1393} (\bibinfo {year} {1994}{\natexlab{b}})}\BibitemShut {NoStop}%
\bibitem [{\citenamefont {Khveshchenko}(1995)}]{khvesh}%
  \BibitemOpen
  \bibfield  {author} {\bibinfo {author} {\bibfnamefont {D.~V.}\ \bibnamefont
  {Khveshchenko}},\ }\bibfield  {title} {\bibinfo {title} {Geometrical approach
  to bosonization of $d>1$ dimensional (non)-fermi liquids},\ }\href
  {https://doi.org/10.1103/PhysRevB.52.4833} {\bibfield  {journal} {\bibinfo
  {journal} {Phys. Rev. B}\ }\textbf {\bibinfo {volume} {52}},\ \bibinfo
  {pages} {4833} (\bibinfo {year} {1995})}\BibitemShut {NoStop}%
\bibitem [{\citenamefont {Houghton}\ \emph {et~al.}(2000)\citenamefont
  {Houghton}, \citenamefont {Kwon},\ and\ \citenamefont
  {Marston}}]{Houghton_2000}%
  \BibitemOpen
  \bibfield  {author} {\bibinfo {author} {\bibfnamefont {A.}~\bibnamefont
  {Houghton}}, \bibinfo {author} {\bibfnamefont {H.-J.}\ \bibnamefont {Kwon}},\
  and\ \bibinfo {author} {\bibfnamefont {J.~B.}\ \bibnamefont {Marston}},\
  }\bibfield  {title} {\bibinfo {title} {Multidimensional bosonization},\
  }\href {https://doi.org/10.1080/000187300243363} {\bibfield  {journal}
  {\bibinfo  {journal} {Advances in Physics}\ }\textbf {\bibinfo {volume}
  {49}},\ \bibinfo {pages} {141} (\bibinfo {year} {2000})}\BibitemShut
  {NoStop}%
\bibitem [{\citenamefont {Delacr\'etaz}\ \emph {et~al.}(2022)\citenamefont
  {Delacr\'etaz}, \citenamefont {Du}, \citenamefont {Mehta},\ and\
  \citenamefont {Son}}]{DDMS2022}%
  \BibitemOpen
  \bibfield  {author} {\bibinfo {author} {\bibfnamefont {L.~V.}\ \bibnamefont
  {Delacr\'etaz}}, \bibinfo {author} {\bibfnamefont {Y.-H.}\ \bibnamefont
  {Du}}, \bibinfo {author} {\bibfnamefont {U.}~\bibnamefont {Mehta}},\ and\
  \bibinfo {author} {\bibfnamefont {D.~T.}\ \bibnamefont {Son}},\ }\bibfield
  {title} {\bibinfo {title} {Nonlinear bosonization of fermi surfaces: The
  method of coadjoint orbits},\ }\href
  {https://doi.org/10.1103/PhysRevResearch.4.033131} {\bibfield  {journal}
  {\bibinfo  {journal} {Phys. Rev. Res.}\ }\textbf {\bibinfo {volume} {4}},\
  \bibinfo {pages} {033131} (\bibinfo {year} {2022})}\BibitemShut {NoStop}%
\bibitem [{\citenamefont {Han}\ \emph {et~al.}(2023)\citenamefont {Han},
  \citenamefont {Desrochers},\ and\ \citenamefont {Kim}}]{kim2023}%
  \BibitemOpen
  \bibfield  {author} {\bibinfo {author} {\bibfnamefont {S.}~\bibnamefont
  {Han}}, \bibinfo {author} {\bibfnamefont {F.}~\bibnamefont {Desrochers}},\
  and\ \bibinfo {author} {\bibfnamefont {Y.~B.}\ \bibnamefont {Kim}},\ }\href
  {https://arxiv.org/abs/2306.14955} {\bibinfo {title} {Bosonization of
  non-fermi liquids}} (\bibinfo {year} {2023}),\ \Eprint
  {https://arxiv.org/abs/2306.14955} {arXiv:2306.14955 [cond-mat.str-el]}
  \BibitemShut {NoStop}%
\bibitem [{\citenamefont {Park}\ and\ \citenamefont
  {Balents}(2024)}]{balents2024}%
  \BibitemOpen
  \bibfield  {author} {\bibinfo {author} {\bibfnamefont {T.}~\bibnamefont
  {Park}}\ and\ \bibinfo {author} {\bibfnamefont {L.}~\bibnamefont {Balents}},\
  }\bibfield  {title} {\bibinfo {title} {{An exact method for bosonizing the
  Fermi surface in arbitrary dimensions}},\ }\href
  {https://doi.org/10.21468/SciPostPhys.16.3.069} {\bibfield  {journal}
  {\bibinfo  {journal} {SciPost Phys.}\ }\textbf {\bibinfo {volume} {16}},\
  \bibinfo {pages} {069} (\bibinfo {year} {2024})}\BibitemShut {NoStop}%
\bibitem [{\citenamefont {Huang}(2024)}]{Huang2024}%
  \BibitemOpen
  \bibfield  {author} {\bibinfo {author} {\bibfnamefont {X.}~\bibnamefont
  {Huang}},\ }\bibfield  {title} {\bibinfo {title} {Effective field theory of
  berry fermi liquid from the coadjoint orbit method},\ }\href
  {https://doi.org/10.1103/PhysRevB.109.235146} {\bibfield  {journal} {\bibinfo
   {journal} {Phys. Rev. B}\ }\textbf {\bibinfo {volume} {109}},\ \bibinfo
  {pages} {235146} (\bibinfo {year} {2024})}\BibitemShut {NoStop}%
\bibitem [{\citenamefont {Ravid}(2024)}]{ravid2024electronslostphasespace}%
  \BibitemOpen
  \bibfield  {author} {\bibinfo {author} {\bibfnamefont {T.}~\bibnamefont
  {Ravid}},\ }\href {https://arxiv.org/abs/2412.00924} {\bibinfo {title}
  {Electrons lost in phase space}} (\bibinfo {year} {2024}),\ \Eprint
  {https://arxiv.org/abs/2412.00924} {arXiv:2412.00924 [cond-mat.str-el]}
  \BibitemShut {NoStop}%
\bibitem [{\citenamefont {Delacrétaz}\ \emph {et~al.}(2025)\citenamefont
  {Delacrétaz}, \citenamefont {Chowdhury},\ and\ \citenamefont
  {Mehta}}]{Delacretaz2025}%
  \BibitemOpen
  \bibfield  {author} {\bibinfo {author} {\bibfnamefont {L.~V.}\ \bibnamefont
  {Delacrétaz}}, \bibinfo {author} {\bibfnamefont {S.~D.}\ \bibnamefont
  {Chowdhury}},\ and\ \bibinfo {author} {\bibfnamefont {U.}~\bibnamefont
  {Mehta}},\ }\href {https://arxiv.org/abs/2501.02073} {\bibinfo {title}
  {Symmetry and causality constraints on fermi liquids}} (\bibinfo {year}
  {2025}),\ \Eprint {https://arxiv.org/abs/2501.02073} {arXiv:2501.02073
  [hep-th]} \BibitemShut {NoStop}%
\bibitem [{\citenamefont {Ye}\ and\ \citenamefont
  {Wang}(2024)}]{ye_wang_first}%
  \BibitemOpen
  \bibfield  {author} {\bibinfo {author} {\bibfnamefont {M.}~\bibnamefont
  {Ye}}\ and\ \bibinfo {author} {\bibfnamefont {Y.}~\bibnamefont {Wang}},\
  }\href {https://arxiv.org/abs/2408.06409} {\bibinfo {title} {Coadjoint-orbit
  effective field theory of a fermi surface in a weak magnetic field}}
  (\bibinfo {year} {2024}),\ \Eprint {https://arxiv.org/abs/2408.06409}
  {arXiv:2408.06409 [cond-mat.str-el]} \BibitemShut {NoStop}%
\bibitem [{\citenamefont {Baym}\ and\ \citenamefont
  {Pethick}(2008)}]{baym2008landau}%
  \BibitemOpen
  \bibfield  {author} {\bibinfo {author} {\bibfnamefont {G.}~\bibnamefont
  {Baym}}\ and\ \bibinfo {author} {\bibfnamefont {C.}~\bibnamefont {Pethick}},\
  }\href@noop {} {\emph {\bibinfo {title} {Landau Fermi-liquid theory: concepts
  and applications}}}\ (\bibinfo  {publisher} {John Wiley \& Sons},\ \bibinfo
  {year} {2008})\BibitemShut {NoStop}%
\bibitem [{\citenamefont {Kohn}(1961)}]{Kohn}%
  \BibitemOpen
  \bibfield  {author} {\bibinfo {author} {\bibfnamefont {W.}~\bibnamefont
  {Kohn}},\ }\bibfield  {title} {\bibinfo {title} {Cyclotron resonance and de
  haas-van alphen oscillations of an interacting electron gas},\ }\href
  {https://doi.org/10.1103/PhysRev.123.1242} {\bibfield  {journal} {\bibinfo
  {journal} {Phys. Rev.}\ }\textbf {\bibinfo {volume} {123}},\ \bibinfo {pages}
  {1242} (\bibinfo {year} {1961})}\BibitemShut {NoStop}%
\bibitem [{\citenamefont {Delacretaz}(2024)}]{Luca_talk}%
  \BibitemOpen
  \bibfield  {author} {\bibinfo {author} {\bibfnamefont {L.}~\bibnamefont
  {Delacretaz}},\ }\bibfield  {title} {\bibinfo {title} {Nonlinear bosonization
  of fermi liquids}} (\bibinfo {year} {2024}),\ \bibinfo {note} {talk at the
  University of Florida.}\BibitemShut {Stop}%
\bibitem [{\citenamefont {Kamenev}(2011)}]{KamenevBook}%
  \BibitemOpen
  \bibfield  {author} {\bibinfo {author} {\bibfnamefont {A.}~\bibnamefont
  {Kamenev}},\ }\href@noop {} {\emph {\bibinfo {title} {Field Theory of
  Non-Equilibrium Systems}}}\ (\bibinfo  {publisher} {Cambridge University
  Press},\ \bibinfo {year} {2011})\BibitemShut {NoStop}%
\bibitem [{\citenamefont {Douglas}\ and\ \citenamefont
  {Nekrasov}(2001)}]{Douglas2001rmp}%
  \BibitemOpen
  \bibfield  {author} {\bibinfo {author} {\bibfnamefont {M.~R.}\ \bibnamefont
  {Douglas}}\ and\ \bibinfo {author} {\bibfnamefont {N.~A.}\ \bibnamefont
  {Nekrasov}},\ }\bibfield  {title} {\bibinfo {title} {Noncommutative field
  theory},\ }\href {https://doi.org/10.1103/RevModPhys.73.977} {\bibfield
  {journal} {\bibinfo  {journal} {Rev. Mod. Phys.}\ }\textbf {\bibinfo {volume}
  {73}},\ \bibinfo {pages} {977} (\bibinfo {year} {2001})}\BibitemShut
  {NoStop}%
\bibitem [{\citenamefont {Floreanini}\ and\ \citenamefont
  {Jackiw}(1987)}]{FJ1987}%
  \BibitemOpen
  \bibfield  {author} {\bibinfo {author} {\bibfnamefont {R.}~\bibnamefont
  {Floreanini}}\ and\ \bibinfo {author} {\bibfnamefont {R.}~\bibnamefont
  {Jackiw}},\ }\bibfield  {title} {\bibinfo {title} {Self-dual fields as
  charge-density solitons},\ }\href
  {https://doi.org/10.1103/PhysRevLett.59.1873} {\bibfield  {journal} {\bibinfo
   {journal} {Phys. Rev. Lett.}\ }\textbf {\bibinfo {volume} {59}},\ \bibinfo
  {pages} {1873} (\bibinfo {year} {1987})}\BibitemShut {NoStop}%
\bibitem [{\citenamefont {{Haldane}}(2018)}]{Haldane2018b}%
  \BibitemOpen
  \bibfield  {author} {\bibinfo {author} {\bibfnamefont {F.~D.~M.}\
  \bibnamefont {{Haldane}}},\ }\bibfield  {title} {\bibinfo {title} {{The
  origin of holomorphic states in Landau levels from non-commutative geometry
  and a new formula for their overlaps on the torus}},\ }\href
  {https://doi.org/10.1063/1.5046122} {\bibfield  {journal} {\bibinfo
  {journal} {Journal of Mathematical Physics}\ }\textbf {\bibinfo {volume}
  {59}},\ \bibinfo {eid} {081901} (\bibinfo {year} {2018})},\ \Eprint
  {https://arxiv.org/abs/1806.10106} {arXiv:1806.10106 [cond-mat.str-el]}
  \BibitemShut {NoStop}%
\bibitem [{\citenamefont {Kohn}(1959)}]{Kohn1959}%
  \BibitemOpen
  \bibfield  {author} {\bibinfo {author} {\bibfnamefont {W.}~\bibnamefont
  {Kohn}},\ }\bibfield  {title} {\bibinfo {title} {Theory of bloch electrons in
  a magnetic field: The effective hamiltonian},\ }\href
  {https://doi.org/10.1103/PhysRev.115.1460} {\bibfield  {journal} {\bibinfo
  {journal} {Phys. Rev.}\ }\textbf {\bibinfo {volume} {115}},\ \bibinfo {pages}
  {1460} (\bibinfo {year} {1959})}\BibitemShut {NoStop}%
\bibitem [{\citenamefont {Mermin}\ and\ \citenamefont
  {Cheng}(1968)}]{mermin-cheng-1968}%
  \BibitemOpen
  \bibfield  {author} {\bibinfo {author} {\bibfnamefont {N.~D.}\ \bibnamefont
  {Mermin}}\ and\ \bibinfo {author} {\bibfnamefont {Y.~C.}\ \bibnamefont
  {Cheng}},\ }\bibfield  {title} {\bibinfo {title} {Fermi-liquid effects in
  magnetoplasma modes in alkali metals},\ }\href
  {https://doi.org/10.1103/PhysRevLett.20.839} {\bibfield  {journal} {\bibinfo
  {journal} {Phys. Rev. Lett.}\ }\textbf {\bibinfo {volume} {20}},\ \bibinfo
  {pages} {839} (\bibinfo {year} {1968})}\BibitemShut {NoStop}%
\bibitem [{\citenamefont {Ledwith}\ \emph {et~al.}(2019)\citenamefont
  {Ledwith}, \citenamefont {Guo}, \citenamefont {Shytov},\ and\ \citenamefont
  {Levitov}}]{levitov-2019}%
  \BibitemOpen
  \bibfield  {author} {\bibinfo {author} {\bibfnamefont {P.}~\bibnamefont
  {Ledwith}}, \bibinfo {author} {\bibfnamefont {H.}~\bibnamefont {Guo}},
  \bibinfo {author} {\bibfnamefont {A.}~\bibnamefont {Shytov}},\ and\ \bibinfo
  {author} {\bibfnamefont {L.}~\bibnamefont {Levitov}},\ }\bibfield  {title}
  {\bibinfo {title} {Tomographic dynamics and scale-dependent viscosity in 2d
  electron systems},\ }\href {https://doi.org/10.1103/PhysRevLett.123.116601}
  {\bibfield  {journal} {\bibinfo  {journal} {Phys. Rev. Lett.}\ }\textbf
  {\bibinfo {volume} {123}},\ \bibinfo {pages} {116601} (\bibinfo {year}
  {2019})}\BibitemShut {NoStop}%
\bibitem [{\citenamefont {{Moiseenko}}\ \emph {et~al.}(2024)\citenamefont
  {{Moiseenko}}, \citenamefont {{M{\"o}nch}}, \citenamefont {{Kapralov}},
  \citenamefont {{Bandurin}}, \citenamefont {{Ganichev}},\ and\ \citenamefont
  {{Svintsov}}}]{2024arXiv240905147M}%
  \BibitemOpen
  \bibfield  {author} {\bibinfo {author} {\bibfnamefont {I.}~\bibnamefont
  {{Moiseenko}}}, \bibinfo {author} {\bibfnamefont {E.}~\bibnamefont
  {{M{\"o}nch}}}, \bibinfo {author} {\bibfnamefont {K.}~\bibnamefont
  {{Kapralov}}}, \bibinfo {author} {\bibfnamefont {D.}~\bibnamefont
  {{Bandurin}}}, \bibinfo {author} {\bibfnamefont {S.}~\bibnamefont
  {{Ganichev}}},\ and\ \bibinfo {author} {\bibfnamefont {D.}~\bibnamefont
  {{Svintsov}}},\ }\bibfield  {title} {\bibinfo {title} {{Testing the
  tomographic Fermi liquid hypothesis with high-order cyclotron resonance}},\
  }\href {https://doi.org/10.48550/arXiv.2409.05147} {\bibfield  {journal}
  {\bibinfo  {journal} {arXiv e-prints}\ ,\ \bibinfo {eid} {arXiv:2409.05147}}
  (\bibinfo {year} {2024})},\ \Eprint {https://arxiv.org/abs/2409.05147}
  {arXiv:2409.05147 [cond-mat.str-el]} \BibitemShut {NoStop}%
\bibitem [{\citenamefont {Altland}\ and\ \citenamefont
  {Simons}(2010)}]{altland}%
  \BibitemOpen
  \bibfield  {author} {\bibinfo {author} {\bibfnamefont {A.}~\bibnamefont
  {Altland}}\ and\ \bibinfo {author} {\bibfnamefont {B.~D.}\ \bibnamefont
  {Simons}},\ }\href@noop {} {\emph {\bibinfo {title} {Condensed Matter Field
  Theory}}},\ \bibinfo {edition} {2nd}\ ed.\ (\bibinfo  {publisher} {Cambridge
  University Press},\ \bibinfo {year} {2010})\BibitemShut {NoStop}%
\bibitem [{\citenamefont
  {Weisstein}(2000{\natexlab{a}})}]{weisstein2000jacobi}%
  \BibitemOpen
  \bibfield  {author} {\bibinfo {author} {\bibfnamefont {E.~W.}\ \bibnamefont
  {Weisstein}},\ }\bibfield  {title} {\bibinfo {title} {Jacobi theta
  functions},\ }\href {https://mathworld.wolfram.com/JacobiThetaFunctions.html}
  {\bibfield  {journal} {\bibinfo  {journal}
  {\url{https://mathworld.wolfram.com/JacobiThetaFunctions.html}}\ } (\bibinfo
  {year} {2000}{\natexlab{a}})}\BibitemShut {NoStop}%
\bibitem [{\citenamefont
  {Weisstein}(2000{\natexlab{b}})}]{weisstein2000riemann}%
  \BibitemOpen
  \bibfield  {author} {\bibinfo {author} {\bibfnamefont {E.~W.}\ \bibnamefont
  {Weisstein}},\ }\bibfield  {title} {\bibinfo {title} {Riemann theta
  function},\ }\href {https://mathworld.wolfram.com/RiemannThetaFunction.html}
  {\bibfield  {journal} {\bibinfo  {journal}
  {\url{https://mathworld.wolfram.com/RiemannThetaFunction.html}}\ } (\bibinfo
  {year} {2000}{\natexlab{b}})}\BibitemShut {NoStop}%
\bibitem [{\citenamefont {Chubukov}\ and\ \citenamefont
  {Maslov}(2012)}]{first-Mats}%
  \BibitemOpen
  \bibfield  {author} {\bibinfo {author} {\bibfnamefont {A.~V.}\ \bibnamefont
  {Chubukov}}\ and\ \bibinfo {author} {\bibfnamefont {D.~L.}\ \bibnamefont
  {Maslov}},\ }\bibfield  {title} {\bibinfo {title} {First-matsubara-frequency
  rule in a fermi liquid. i. fermionic self-energy},\ }\href
  {https://doi.org/10.1103/PhysRevB.86.155136} {\bibfield  {journal} {\bibinfo
  {journal} {Phys. Rev. B}\ }\textbf {\bibinfo {volume} {86}},\ \bibinfo
  {pages} {155136} (\bibinfo {year} {2012})}\BibitemShut {NoStop}%
\bibitem [{\citenamefont {Wang}()}]{wang_unpublished}%
  \BibitemOpen
  \bibfield  {author} {\bibinfo {author} {\bibfnamefont {Y.}~\bibnamefont
  {Wang}},\ }\href@noop {} {}\bibinfo {howpublished} {Unpublished}\BibitemShut
  {NoStop}%
\bibitem [{\citenamefont {Giamarchi}\ and\ \citenamefont
  {Schulz}(1988)}]{giamarchiSchulz1988}%
  \BibitemOpen
  \bibfield  {author} {\bibinfo {author} {\bibfnamefont {T.}~\bibnamefont
  {Giamarchi}}\ and\ \bibinfo {author} {\bibfnamefont {H.~J.}\ \bibnamefont
  {Schulz}},\ }\bibfield  {title} {\bibinfo {title} {Anderson localization and
  interactions in one-dimensional metals},\ }\href
  {https://doi.org/10.1103/PhysRevB.37.325} {\bibfield  {journal} {\bibinfo
  {journal} {Phys. Rev. B}\ }\textbf {\bibinfo {volume} {37}},\ \bibinfo
  {pages} {325} (\bibinfo {year} {1988})}\BibitemShut {NoStop}%
\bibitem [{\citenamefont {Gornyi}\ \emph {et~al.}(2007)\citenamefont {Gornyi},
  \citenamefont {Mirlin},\ and\ \citenamefont
  {Polyakov}}]{GornyiMirlinPolyakov2007}%
  \BibitemOpen
  \bibfield  {author} {\bibinfo {author} {\bibfnamefont {I.~V.}\ \bibnamefont
  {Gornyi}}, \bibinfo {author} {\bibfnamefont {A.~D.}\ \bibnamefont {Mirlin}},\
  and\ \bibinfo {author} {\bibfnamefont {D.~G.}\ \bibnamefont {Polyakov}},\
  }\bibfield  {title} {\bibinfo {title} {Electron transport in a disordered
  luttinger liquid},\ }\href {https://doi.org/10.1103/PhysRevB.75.085421}
  {\bibfield  {journal} {\bibinfo  {journal} {Phys. Rev. B}\ }\textbf {\bibinfo
  {volume} {75}},\ \bibinfo {pages} {085421} (\bibinfo {year}
  {2007})}\BibitemShut {NoStop}%
\bibitem [{\citenamefont {Dingle}\ and\ \citenamefont {Bragg}(1952)}]{Dingle}%
  \BibitemOpen
  \bibfield  {author} {\bibinfo {author} {\bibfnamefont {R.~B.}\ \bibnamefont
  {Dingle}}\ and\ \bibinfo {author} {\bibfnamefont {W.~L.}\ \bibnamefont
  {Bragg}},\ }\bibfield  {title} {\bibinfo {title} {Some magnetic properties of
  metals ii. the influence of collisions on the magnetic behaviour of large
  systems},\ }\href {https://doi.org/10.1098/rspa.1952.0056} {\bibfield
  {journal} {\bibinfo  {journal} {Proceedings of the Royal Society of London.
  Series A. Mathematical and Physical Sciences}\ }\textbf {\bibinfo {volume}
  {211}},\ \bibinfo {pages} {517} (\bibinfo {year} {1952})},\ \Eprint
  {https://arxiv.org/abs/https://royalsocietypublishing.org/doi/pdf/10.1098/rspa.1952.0056}
  {https://royalsocietypublishing.org/doi/pdf/10.1098/rspa.1952.0056}
  \BibitemShut {NoStop}%
\bibitem [{\citenamefont {Brailsford}(1966)}]{Brailsford}%
  \BibitemOpen
  \bibfield  {author} {\bibinfo {author} {\bibfnamefont {A.~D.}\ \bibnamefont
  {Brailsford}},\ }\bibfield  {title} {\bibinfo {title} {Influence of
  impurities on the de haas-van alphen effect},\ }\href
  {https://doi.org/10.1103/PhysRev.149.456} {\bibfield  {journal} {\bibinfo
  {journal} {Phys. Rev.}\ }\textbf {\bibinfo {volume} {149}},\ \bibinfo {pages}
  {456} (\bibinfo {year} {1966})}\BibitemShut {NoStop}%
\bibitem [{\citenamefont {Adamov}\ \emph {et~al.}(2006)\citenamefont {Adamov},
  \citenamefont {Gornyi},\ and\ \citenamefont
  {Mirlin}}]{AdamovGornyiMirlin2006}%
  \BibitemOpen
  \bibfield  {author} {\bibinfo {author} {\bibfnamefont {Y.}~\bibnamefont
  {Adamov}}, \bibinfo {author} {\bibfnamefont {I.~V.}\ \bibnamefont {Gornyi}},\
  and\ \bibinfo {author} {\bibfnamefont {A.~D.}\ \bibnamefont {Mirlin}},\
  }\bibfield  {title} {\bibinfo {title} {Interaction effects on
  magneto-oscillations in a two-dimensional electron gas},\ }\href
  {https://doi.org/10.1103/PhysRevB.73.045426} {\bibfield  {journal} {\bibinfo
  {journal} {Phys. Rev. B}\ }\textbf {\bibinfo {volume} {73}},\ \bibinfo
  {pages} {045426} (\bibinfo {year} {2006})}\BibitemShut {NoStop}%
\end{thebibliography}%

\end{document}